\documentclass[11pt]{article}
\usepackage[bookmarks,bookmarksnumbered]{hyperref}
\usepackage{amsmath,XoohmE,multirow}
\usepackage{graphicx,color}
\usepackage{longtable}
\SfTitles 
\seceq 
\newcounter{inp}
\newenvironment{inp}[1][]{\refstepcounter{inp}\par\noindent{\em\underline{\textbf{Input~(\theinp#1)}}}\nobreak\vglue0pt\noindent}{}
\def\outp{\medskip\noindent{\em\underline{\textbf{Output~(\theinp)}}}\nobreak\vglue0pt\noindent}
\def\binp{\begin{inp}}
\def\einp{\end{inp}\par}
\def\[#1 {\hglue#1mm\ignorespaces}
\def\Fl#1{\texttt{\itshape#1}}
\def\fg{\mathfrak{g}}

\def\piC#1#2{\begin{picture}(0,0)\put(#1){#2}\end{picture}}

\definecolor{gReen}{rgb}{.90,1,.90}
\definecolor{Green}{rgb}{0,.5,0}

\allowdisplaybreaks
\begin{document}
\pagestyle{empty}
\begingroup
\renewcommand{\thefootnote}{\fnsymbol{footnote}}
\begin{center}
{\LARGE\sf\bfseries\boldmath
 On $\ZZ_N$-Invariant Subgroups of Semi-Simple Lie Groups}\\[3mm]
{\sf\bfseries
  M.K.~Ahsan\footnote{Email: mohammad.ahsan@utdallas.edu}
  \,{\rm and}
  T.~H\"{u}bsch\ft{Email: thubsch@howard.edu}
}\\[1mm]
{\small\it
  $^*$\,Department of Mathematical Sciences,
      University of Texas at Dallas, Richardson TX 75080.
}\\
{\small\it
  $^\dag$\,Department of Physics and Astronomy,
      Howard University, Washington, DC 20059
}\\[15mm]
{\bf ABSTRACT}\\[5mm]
\parbox{15cm}{We employ \textsl{Mathematica} to find $\ZZ_N$-invariant subgroups of $E_8$ for application in $M$-theory. These $\ZZ_N$-invariant subgroups are phenomenologically important and in some cases they resemble the gauge groups of our real world. We present a specific example of $\ZZ_7$-invariant subgroups of $E_8$, which turn up in orbifold compactification of $M$-theory. Moreover, the procedure can be applied for any $\ZZ_N$ group that acts by shifts (translations) in the root lattice of semisimple Lie groups with $A_n,B_n,C_n,D_n,E_6,E_7$ and $E_8$ factors.}
\end{center}
\noindent
 PACS: 02.20.Rt, 11.25.Mj\hfill
 Keywords:~\parbox[t]{45mm}{orbifold compactification,\\[-1mm] string-theory, $M$-theory}
\endgroup
\vfill
\pagestyle{plain}
\setcounter{footnote}{0}
\setcounter{page}{1}
\section{Introduction}
\label{intro}
In models where part of spacetime is compactified, the geometry of compact space affects the gauge symmetries of the model. Herein, we consider the Ho\v{r}ava-Witten $M$-theory\cite{rPHEW1,rBBS}, where the 11$^\text{th}$ dimension is compactified to an interval, $I$, and there are two ten-dimensional planes at the boundaries of $I$. It is convenient to identify $I=S^1/\ZZ_2$, acting as $\ZZ_2:\f\to-\f$, so that the boundary of $I$ consists of the fixed points of this $\ZZ_2$-action. On each one of these ten-dimensional spacetime planes there is an independent copy of $E_8$ gauge fields (principal vector bundle). To produce considerably more realistic models with 4-dimensional spacetime, one may proceed as follows:
\begin{enumerate}\itemsep=-3pt\vspace{-2mm}
 \item Impose twisted periodicity conditions on six of the ten dimensions of the boundary spacetime planes, passing $\IR^6\to(T^6/\D)=((\IR^6/\L)/\D)$, where $\L$ is a suitable 6-dimensional lattice and $\D$ is a symmetry of $\L$. We consider $\D=\ZZ_N$.
 \item Simultaneously embed the $\D$ action into the $E_8$ structure group of the gauge fields on each of the two boundary-spacetimes, the structure groups are broken to subgroups of $E_8$ that are invariant with respect to the $\D$-action.
\end{enumerate}\vspace{-2mm}
This is referred to as ``compactifying the Ho\v{r}ava-Witten $M$-theory on a $T^6/\D$ orbifold'', and $\D$ is the ``orbifold group.'' Typically, $\D$ acts by rotations on the compact space coordinates, and at the same time by shifts (translations) in the $E_8$ root lattice\cite{rKP-Anom,rHM-112}.

In Ref.\cite{rAHZ7}, we have constructed $\ZZ_7$-orbifold models in $M$-theory. We used \textsl{Mathematica} to find the $\ZZ_7$-invariant subgroups of $E_8$. In this paper we present the details of the \textsl{Mathematica} computation codes and the procedure that we have used in Ref.\cite{rAHZ7}. This procedure may be used, perhaps with minor adaptations, for higher order (iterated) orbifolds as well, and in situations where one needs to find the $\ZZ_N$-invariant subgroups of any of the semisimple Lie groups with $A_n,B_n,C_n,D_n,E_6,E_7$ and $E_8$ factors, where $\ZZ_N$ acts by shifts in the root lattice.
\section{The Algorithm}
\label{algo}
Consider the root lattice $\bm{\cal P}$ of one of the simple Lie algebras $\mathfrak{g}=A_n, B_n, C_n, D_n, E_6, E_7$ or $E_8$.
Let $\bm{v}$ denote a shift (translation) vector in $\bm{\cal P}$ acting as $e^{2\p i \bm{v}{\cdot}\bm{p}}\ket{\bm{p}}$ on $\ket{\bm{p}}\in\bm{\cal P}$\cite{rKP-Anom,rHM-112}; require moreover that $(e^{2\p i \bm{v}{\cdot}\bm{p}})^N=\Ione$, so that $\bm{v}$ generates a $\ZZ_N$ action on $\bm{\cal P}$, and thus on $G$.
 The root vectors of $\fg$ that are invariant with respect to this $\bm{v}$-action
\begin{equation}
 e^{2\p i \bm{v}{\cdot}\bm{p}}\ket{\bm{p}} = \ket{\bm{p}}, \qquad
 \bm{p}\in\bm{\cal P},
\label{shift}
\end{equation}
are the root vectors of a subgroup $H\subset G$ that is invariant with respect to the $\ZZ_N$-action generated by $\bm{v}$. Different shift vectors $\bm{v}$ define different $\ZZ_N$-actions, and therefore different $\ZZ_N$-invariant subgroups of $G$. Upon identifying those that are equivalent by $G$-conjugation, we find the inequivalent $\ZZ_N$-invariant subgroups, $H_I$, for $I=1,2,\ldots$ Without loss of generality, we restrict the $[\frac{k}N\>(\text{mod}\>1)]$-valued components of $\bm{v}$ in\eq{shift} to the standard range $\big\{0,\frac1N,\frac2N,\cdots,\frac{N-1}N\big\}$.

\paragraph{\bfseries Note:}
As the so-defined $\ZZ_N$-invariant subgroups $H_I\subset G$ are explicitly defined in terms of the root lattice of $G$, they are by definition {\em\/regular\/}\cite{rD-LieSub,rNJ-LA,rJH-ILieA+RT,rG-LieG,rWyb,rSsky,rCahn}. In addition, the condition\eq{shift} is trivially satisfied for the zero-weight vectors $\bm{p}_{\sss C}=\texttt{\{0,0,$\cdots$,0\}}$ corresponding to Cartan generators of $G$. Therefore,
\begin{equation}
   \rk(H_I)=\rk(G),
 \label{e:rkHI}
\end{equation}
and all so-defined $\ZZ_N$-invariant regular subgroups of $G$ also have maximal rank.

\begin{description}

\item[Step\,1:]  Find the set of positive root vectors of $G$, denoted $\bm{\cal W}$.

\item[Step\,2:] Based on the above restrictions, we construct all possible $\ZZ_N$ shift vectors $\bm{v}$.

\item[Step\,3:] Find all the subgroups\ft{For all simple Lie groups of rank $\leq8$ and several of higher rank, the maximal subgroups are listed in the literature\cite{rD-LieSub,rSsky,rMcKP}; for the general procedure, see Ref.\cite{rD-LieSub} and also Appendix~\ref{s:A}.} $H_I\subset G$.

\item[Step\,4:] For each one of the subgroups $H_I\subset G$, define the following four variables:

 \item[\hspace{2pc}] $t$ is the set of positive root vectors of $H_I\subset G$;\vspace{-2mm}

 \item[\hspace{2pc}] $p\Defl|t|$ is total number of positive root vectors in $H_I\subset G$;\vspace{-2mm}

 \item[\hspace{2pc}] $r\Defl\rk'(H_I)$, {\em\/defined\/} as the rank of semisimple part of $H_I\subset G$, \ie, {\em\/without\/} $U(1)$-factors;\vspace{-2mm}

 \item[\hspace{2pc}] $m$ is number of $A_1$ factors, if any, in $H_I\subset G$.

These three variables can be read off by looking at the  subgroup $H_I$ and can be used as identifiers of the group. If these three variables do not suffice to identify the Dynkin type of $H_I\subset G$, define another variable:\vspace{-2mm}

 \item[\hspace{2pc}] $m_2$ is the number of $A_2$ factors in $H_I$, if any.
 
If $\{p,r,m,m_2\}$ turns out not to suffice to identify $H_I\subset G$ unambiguously, we look for $A_3$, $A_4$\dots\ factors in $H_I$, the numbers of which, $m_3$, $m_4$\dots, will be necessary to identify $H_I\subset G$ unambiguously.

\item[Step\,5:] Pick the first $\bm{v}$ from {\bf Step\,2}.\vspace{-2mm}
 \begin{description}

  \item[Step\,5.a:] Set $t=\varnothing$.
  For all $\bm{w}_a\in\bm{\cal W}$, if\ft{Since the root lattice shift $\bm{v}$ corresponds to a generator $g(\bm{v})\in\ZZ_N$ so that all elements of $\ZZ_N$ are powers of $g(\bm{v})$, it follows that root vectors satisfying $\bm{v}{\cdot}\bm{w}_a=\ZZ$ are in fact invariant with respect to all of $\ZZ_N$.} $\bm{v}{\cdot}\bm{w}_a=\ZZ$, append $\bm{w}_a$ into the set $t$.

  \item[Step\,5.b:] Compute $\{p,r,m,\dots\}$ of this $t$ (see Section~\ref{znisg} for the procedure).

  \item[Step\,5.c:] Identify the subgroup $H_I\subset G$ by comparing $\{p,r,m,\dots\}$  with the list from {\bf Step\,4}.
 \end{description}

\item[Step\,6:] Pick the next $\bm{v}$ from {\bf Step\,2}, and go to {\bf Step\,5.a}.
\end{description}

Steps\,1--4 are preparatory. In particular, Step\,4 sets up the string of identifiers $\{p,r,m,m_2,\dots\}$ as an ``address'' of the regular subgroups $H_I$ ($I=1,2,3,\ldots$) of a given simple Lie group $G$. For the purposes of specific applications, such as in $M$-theory\cite{rBBS,rAHZ7,rKKKO} with $G=E_8$ and $\ZZ_N$ acting by translations in the root lattice\eq{shift}, a subset of the identifiers $\{p,r,m,m_2,\dots\}$ sufficed.

\section{Roots and Shift Vectors}
\label{rsv}
\begin{figure}[ht]
 \unitlength=.4mm
$$
\begin{picture}(138,108)(0,-84)
\put(14,0){\circle{5}}
\put(29,0){\circle{5}}
\put(44,0){\circle{5}}
\put(59,0){\circle{5}}
\put(74,0){\circle{5}}
\put(89,0){\circle{5}}
\put(74,20){\circle{5}}
\put(104,0){\circle{5}}
\put(16,0){\line(1,0){11}}
\put(31,0){\line(1,0){11}}
\put(46,0){\line(1,0){11}}
\put(61,0){\line(1,0){11}}
\put(91,0){\line(1,0){11}}
\put(76,0){\line(1,0){11}}
\put(74,2){\line(0,1){16}}
\put(13,-8){$\SSS\a_1$}
\put(28,-8){$\SSS\a_2$}
\put(43,-8){$\SSS\a_3$}
\put(58,-8){$\SSS\a_4$}
\put(73,-8){$\SSS\a_5$}
\put(88,-8){$\SSS\a_6$}
\put(103,-8){$\SSS\a_7$}
\put(78,15){$\SSS\a_8$}
\put(-2,18){\small E.B.~Dynkin\cite{rD-LieSub,rG-LieG}}
\put(6,-24){\small Adjoint: \texttt{\{1,0,0,0,0,0,0,0\}}}
\put(12,-84){\small$\left[\begin{matrix}
                                 2&\bar1&0&0&0&0&0&0\\
                                 \bar1&2&\bar1&0&0&0&0&0\\
                                 0&\bar1&2&\bar1&0&0&0&0\\
                                 0&0&\bar1&2&\bar1&0&0&0\\
                                 0&0&0&\bar1&2&\bar1&0&\bar1\\
                                 0&0&0&0&\bar1&2&\bar1&0\\
                                 0&0&0&0&0&\bar1&2&0\\
                                 0&0&0&0&\bar1&0&0&2\\
                          \end{matrix}\right]$}
\end{picture}
\begin{picture}(138,108)(0,-84)
\put(14,0){\circle{5}}
\put(29,0){\circle{5}}
\put(44,0){\circle{5}}
\put(59,0){\circle{5}}
\put(74,0){\circle{5}}
\put(89,0){\circle{5}}
\put(44,20){\circle{5}}
\put(104,0){\circle{5}}
\put(16,0){\line(1,0){11}}
\put(31,0){\line(1,0){11}}
\put(46,0){\line(1,0){11}}
\put(61,0){\line(1,0){11}}
\put(91,0){\line(1,0){11}}
\put(76,0){\line(1,0){11}}
\put(44,2){\line(0,1){16}}
\put(13,-8){$\SSS\a_1$}
\put(28,-8){$\SSS\a_3$}
\put(43,-8){$\SSS\a_4$}
\put(58,-8){$\SSS\a_5$}
\put(73,-8){$\SSS\a_6$}
\put(88,-8){$\SSS\a_7$}
\put(103,-8){$\SSS\a_8$}
\put(32,15){$\SSS\a_2$}
\put(54,18){\small J.~Hymphreys\cite{rJH-ILieA+RT}}
\put(6,-24){\small Adjoint: \texttt{\{0,0,0,0,0,0,0,1\}}}
\put(12,-84){\small$\left[\begin{matrix}
                                 2&0&\bar1&0&0&0&0&0\\
                                 0&2&0&\bar1&0&0&0&0\\
                                 \bar1&0&2&\bar1&0&0&0&0\\
                                 0&\bar1&\bar1&2&\bar1&0&0&0\\
                                 0&0&0&\bar1&2&\bar1&0&0\\
                                 0&0&0&0&\bar1&2&\bar1&0\\
                                 0&0&0&0&0&\bar1&2&\bar1\\
                                 0&0&0&0&0&0&\bar1&2\\
                           \end{matrix}\right]$}
\end{picture}
\begin{picture}(138,108)(0,-84)
\put(14,0){\circle{5}}
\put(29,0){\circle{5}}
\put(44,0){\circle{5}}
\put(59,0){\circle{5}}
\put(74,0){\circle{5}}
\put(89,0){\circle{5}}
\put(44,20){\circle{5}}
\put(104,0){\circle{5}}
\put(16,0){\line(1,0){11}}
\put(31,0){\line(1,0){11}}
\put(46,0){\line(1,0){11}}
\put(61,0){\line(1,0){11}}
\put(91,0){\line(1,0){11}}
\put(76,0){\line(1,0){11}}
\put(44,2){\line(0,1){16}}
\put(13,-8){$\SSS\a_1$}
\put(28,-8){$\SSS\a_2$}
\put(43,-8){$\SSS\a_3$}
\put(58,-8){$\SSS\a_4$}
\put(73,-8){$\SSS\a_5$}
\put(88,-8){$\SSS\a_6$}
\put(103,-8){$\SSS\a_7$}
\put(32,15){$\SSS\a_8$}
\put(54,18){\small B.~Wybourne\cite{rWyb,rSsky,rCahn}}
\put(6,-24){\small Adjoint: \texttt{\{0,0,0,0,0,0,1,0\}}}
\put(12,-84){\small$\left[\begin{matrix}
                                 2&\bar1&0&0&0&0&0&0\\
                                 \bar1&2&\bar1&0&0&0&0&0\\
                                 0&\bar1&2&\bar1&0&0&0&\bar1\\
                                 0&0&\bar1&2&\bar1&0&0&0\\
                                 0&0&0&\bar1&2&\bar1&0&0\\
                                 0&0&0&0&\bar1&2&\bar1&0\\
                                 0&0&0&0&0&\bar1&2&0\\
                                 0&0&\bar1&0&0&0&0&2\\
                          \end{matrix}\right]$}
\end{picture}
$$
\vspace{5mm}
\caption[Dynkin diagram, highest root and Cartan matrix of $E_8$]{The Dynkin diagram, the highest root of the adjoint representation and the Cartan matrix of $E_8$, given in three fairly standard conventions and some corresponding references.}
\label{f:3As}
\end{figure}
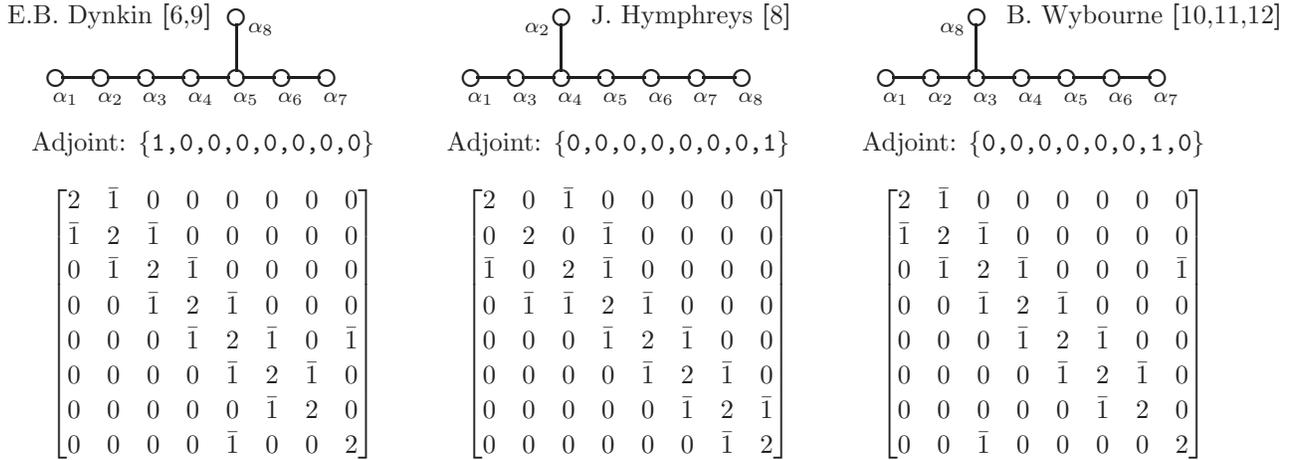

We take the adjoint representation of the group $G$ and calculate its positive root vectors from the highest root using the standard algorithm\cite{rD-LieSub,rJH-ILieA+RT,rG-LieG,rWyb,rSsky,rCahn}. Take for example the group $G=E_8$.
Any concrete representation of these roots will depend on a choice of a basis, and there exist at least three fairly standard conventions, corresponding to the labeling of nodes of the Dynkin diagram of $E_8$, as shown\ft{To save space, negative root vector components are denoted by an over-bar: $\texttt{\=1}\define-1$, $\texttt{\=2}\define-2$, \etc} in Figure~\ref{f:3As}.
Being interested primarily in high energy physics applications such as in Ref.\cite{rBBS,rAHZ7,rKKKO}, we follow the conventions of Refs.\cite{rWyb,rSsky}, which provide the decades-long standard in the high energy physics.

The highest root of the irreducible (248-dimensional) adjoint representation of $E_8$ is {\tt \{0,0,0,0,0,0,1,0\}}.
 The entire root system can be obtained from the highest root by subtracting from it the positive simple root vectors as follows: in any given root vector $\bm{w}$, a positive value of the $n^\text{th}$ component, $\bm{w}$\texttt{[n]}, indicates
 the number of times the $n^\text{th}$ positive simple root $\bm\a_n$ can be subtracted from $\bm{w}$ minus the number of times $\bm\a_n$ can be added to $\bm{w}$ so as
to get another root or zero\cite{rWyb,rCahn}.
 For example, $\bm\a_1=\texttt{\{2,\=1,0,0,0,0,0,0\}}$ is the first positive simple root (and the $1^\text{st}$ row in the Cartan matrix; see Figure~\ref{f:3As}); it may be subtracted from itself twice\ft{To be meticulous, the fact that the first component of $\bm\a_1=\texttt{\{2,\=1,0,0,0,0,0,0\}}$ is $\bm\a_1\texttt{[1]}=+2$ merely means that we can subtract $\bm\a_1$ from itself two more times than we can add $\bm\a_1$ to itself, and still get a root or zero. However, since $\bm\a_1{+}\bm\a_1\neq0$ can be shown not to be a root, it follows that $\bm\a_1$ can be added to itself zero number of times while staying in the root system, and so can be subtracted from itself precisely two times.}, producing:
\begin{equation}
 \texttt{\{2,\=1,0,0,0,0,0,0\}} \tooo{-\bm\a_1~}
 \texttt{\{0,0,0,0,0,0,0,0\}} \tooo{-\bm\a_1~}
 \texttt{\{\=2,1,0,0,0,0,0,0\}}.
\end{equation}
All three of these vectors are indeed in the root system of $E_8$. Starting with $\bm\l=\texttt{\{0,0,0,0,0,0,1,0\}}$, the positive simple root $\bm\a_7=\texttt{\{0,0,0,0,0,\=1,2,0\}}$ may be subtracted once (since $\bm\l$ is the highest root, no positive root can be added and still get a root):
\begin{alignat}9
 \texttt{\{0,0,0,0,0,0,1,0\}} &\tooo{-\bm\a_7~}
 \texttt{\{0,0,0,0,0,1,\=1,0\}},
\intertext{whereupon $\bm\a_6=\texttt{\{0,0,0,0,\=1,2,\=1,0\}}$ may be subtracted once:}
 \texttt{\{0,0,0,0,0,1,\=1,0\}} &\tooo{-\bm\a_6~}
 \texttt{\{0,0,0,0,1,\=1,0,0\}},\qquad \etc
\end{alignat}
Proceeding in this way halts with \texttt{\{0,0,0,0,0,0,\=1,0\}}, having produced 240 (nonzero) root vectors and eight copies of \texttt{\{0,0,0,0,0,0,0,0\}}. Jointly, they span the 248-dimensional adjoint representation of $E_8$.

In fact, every finite-dimensional unitary representation of any semisimple Lie group may be represented in a similar way: We recall that all such representations are spanned by {\em\/weight\/} vectors that are determined by a {\em\/highest weight\/} from which all others are obtained by iteratively subtracting the positive simple roots as outlined above; see Refs.\cite{rWyb,rSsky,rCahn,rJH-ILieA+RT}.
 By definition of $\bm\l$ being the highest weight, no positive simple root may be added to it and get a vector within the weight system of $\bm\l$. Therefore, a positive $n^\text{th}$ component $\bm\l\texttt{[n]}>0$ in the highest weight $\bm\l$ necessarily means that $\bm\a_n$ may be subtracted $\bm\l\texttt{[n]}>0$ number of times from $\bm\l$; plot the so-obtained ``$\bm\a_n$-descendants of $\bm\l$,'' $(\bm\l{-}k\bm\a_n)$, $k$ levels below $\bm\l$. Now proceed downward level by level, seeking an $m^\text{th}\neq n^\text{th}$ positive component in an $\bm\a_n$-descendant weight of $\bm\l$, from which to construct $\bm\a_m$-decendants. Starting from a level where the $m^\text{th}\neq n^\text{th}$ component $(\bm\l{-}k\bm\a_n)\texttt{[m]}>0$ but $(\bm\l{-}k\bm\a_n{+}\bm\a_m)$ is not in the weight-system (immediately above $(\bm\l{-}k\bm\a_n)$) implies that $\bm\a_m$ can be subtracted from $(\bm\l{-}k\bm\a_n)$ precisely $(\bm\l{-}k\bm\a_n)$\texttt{[m]} number of times, producing a chain of $\bm\a_m$-descendants. Proceeding in this fashion eventually terminates and generates the complete weight system when starting from highest weights $\bm\l$ that define finite-dimensional representations\cite{rNJ-LA,rJH-ILieA+RT,rWyb,rSsky,rCahn}.
 Of course, one can just as easily start from the {\em\/lowest weight\/} and add the positive simple roots in the analogous fashion. In the special case of the adjoint representation, which is our focus at present, the nonzero weight vectors are called {\em\/root\/} vectors instead.

By plotting the weights (roots) below those from which they are obtained by subtracting positive simple roots and connecting them by arrows (for illustration, see\eq{e:B2} below), we obtain a  ``spindle shaped'' graph called the weight (root) diagram of the (adjoint) representation.
 In the root diagram of the adjoint representation of a group $G$ of rank $r$, the middle row of the root diagram is populated by $r$ copies of $\texttt{\{0,\ldots,0\}}$, representing the $r$ Cartan generators. The row immediately above the middle is populated by the $r$ positive simple root vectors; the roots above the middle row are the positive root vectors of $G$, while the roots below the middle row are the negative root vectors and are the sign-reversed copies of the positive root vectors. For every Lie group and its algebra, it therefore suffices to map out the subsystem of positive roots.
 
\paragraph{\boldmath$E_8$:}
The following \textsl{Mathematica} code computes the 120 positive root vectors of $E_8$ following the conventions of Ref.\cite{rWyb,rSsky}. The code is adapted to any other convention by changing the basis for both the Cartan matrix and the highest root, \ie, the \textsl{Mathematica} variables \texttt{a} and \texttt{g[0]}, respectively; for those displayed in Figure~\ref{f:3As}, a simple permutation of columns and rows will suffice. Also, we use ``external/global'' variable arrays so that the intermediate computations are all accessible, \eg, for troubleshooting and for tracing the functioning of the code; it is then necessary to start with clearing the required symbols, listed explicitly for each code. The Reader may also find the global command \texttt{ClearAll["Global`*"]} useful, which clears all user-defined variables from previous computations.

\binp
{\tt
ClearAll[a, g, e];\\
 a = \{\{2, \=1, 0, 0, 0, 0, 0, 0\},\\*\hglue10mm
       \{\=1, 2, \=1, 0, 0, 0, 0, 0\},\\*\hglue10mm
       \{0, \=1, 2, \=1, 0, 0, 0, \=1\},\\*\hglue10mm
       \{0, 0, \=1, 2, \=1, 0, 0, 0\},\\*\hglue10mm
       \{0, 0, 0, \=1, 2, \=1, 0, 0\},\\*\hglue10mm
       \{0, 0, 0, 0, \=1, 2, \=1, 0\},\\*\hglue10mm
       \{0, 0, 0, 0, 0, \=1, 2, 0\},\\*\hglue10mm
       \{0, 0, \=1, 0, 0, 0, 0, 2\}\};\\
g[0] = \{\{0, 0, 0, 0, 0, 0, 1, 0\}\};\\
g[1] = Table[
   Flatten[g[0]] - a[[Flatten[Position[Flatten[g[0]], 1]][[p]]]],\\ 
   \hspace*{1.4cm}\{p, Length[Flatten[Position[Flatten[g[0]], 1]]]\}];\\
e[x\_] := e[x] = 
   Union[Flatten[\{Table[
       Table[If[g[x][[j]][[i]] == 1, g[x][[j]] - a[[i]],\\ 
         \hspace*{1.4cm}If[g[x][[j]][[i]] == 2, g[x][[j]] - a[[i]]]], \{i, 8\}], \{j, 
        Length[g[x]]\}],\\ 
      \hspace*{1.4cm}Table[Table[
        If[g[x - 1][[l]][[k]] == 2, g[x - 1][[l]] - 2 a[[k]]], \{k, 
         8\}],\\\hspace*{1.2cm} \{l, Length[g[x - 1]]\}]\}, 2]];\\
g[x\_] := g[x] = 
  If[MemberQ[e[x - 1], Null], Delete[e[x - 1], 1], e[x - 1]];\\
Flatten[Table[g[m], \{m, 0, 28\}], 1]}
\label{c:E8}

\outp
{\tt
\{\{0,0,0,0,0,0,1,0\},\{0,0,0,0,0,1,\=1,0\},%
\{0,0,0,0,1,\=1,0,0\},\{0,0,0,1,\=1,0,0,0\},\{0,0,1,\=1,0,0,0,0\},\\$~$
\{0,1,\=1,0,0,0,0,1\},\{0,1,0,0,0,0,0,\=1\},%
\{1,\=1,0,0,0,0,0,1\},\{\=1,0,0,0,0,0,0,1\},\{1,\=1,1,0,0,0,0,\=1\},\\$~$
\{\=1,0,1,0,0,0,0,\=1\},\{1,0,\=1,1,0,0,0,0\},%
\{\=1,1,\=1,1,0,0,0,0\},\{1,0,0,\=1,1,0,0,0\},\{\=1,1,0,\=1,1,0,0,0\},\\$~$
\{0,\=1,0,1,0,0,0,0\},\{1,0,0,0,\=1,1,0,0\},\{\=1,1,0,0,\=1,1,0,0\},%
\{0,\=1,1,\=1,1,0,0,0\},\{1,0,0,0,0,\=1,1,0\},\\$~$
\{\=1,1,0,0,0,\=1,1,0\},\{0,\=1,1,0,\=1,1,0,0\},%
\{0,0,\=1,0,1,0,0,1\},\{1,0,0,0,0,0,\=1,0\},\{\=1,1,0,0,0,0,\=1,0\},\\$~$
\{0,\=1,1,0,0,\=1,1,0\},\{0,0,\=1,1,\=1,1,0,1\},%
\{0,0,0,0,1,0,0,\=1\},\{0,\=1,1,0,0,0,\=1,0\},\{0,0,\=1,1,0,\=1,1,1\},\\$~$
\{0,0,0,\=1,0,1,0,1\},\{0,0,0,1,\=1,1,0,\=1\},%
\{0,0,\=1,1,0,0,\=1,1\},\{0,0,0,\=1,1,\=1,1,1\},\{0,0,0,1,0,\=1,1,\=1\},\\$~$
\{0,0,1,\=1,0,1,0,\=1\},\{0,0,0,\=1,1,0,\=1,1\},%
\{0,0,0,0,\=1,0,1,1\},\{0,0,0,1,0,0,\=1,\=1\},\{0,0,1,\=1,1,\=1,1,\=1\},\\$~$
\{0,1,\=1,0,0,1,0,0\},\{0,0,0,0,\=1,1,\=1,1\},%
\{0,0,1,\=1,1,0,\=1,\=1\},\{0,0,1,0,\=1,0,1,\=1\},\{0,1,\=1,0,1,\=1,1,0\},\\$~$
\{1,\=1,0,0,0,1,0,0\},\{\=1,0,0,0,0,1,0,0\},%
\{0,0,0,0,0,\=1,0,1\},\{0,0,1,0,\=1,1,\=1,\=1\},\{0,1,\=1,0,1,0,\=1,0\},\\$~$
\{0,1,\=1,1,\=1,0,1,0\},\{1,\=1,0,0,1,\=1,1,0\},%
\{\=1,0,0,0,1,\=1,1,0\},\{0,0,1,0,0,\=1,0,\=1\},\{0,1,\=1,1,\=1,1,\=1,0\},\\$~$
\{0,1,0,\=1,0,0,1,0\},\{1,\=1,0,0,1,0,\=1,0\},%
\{1,\=1,0,1,\=1,0,1,0\},\{\=1,0,0,0,1,0,\=1,0\},\{\=1,0,0,1,\=1,0,1,0\},\\$~$
\{0,1,\=1,1,0,\=1,0,0\},\{0,1,0,\=1,0,1,\=1,0\},%
\{1,\=1,0,1,\=1,1,\=1,0\},\{1,\=1,1,\=1,0,0,1,0\},\{\=1,0,0,1,\=1,1,\=1,0\},\\$~$
\{\=1,0,1,\=1,0,0,1,0\},\{0,1,0,\=1,1,\=1,0,0\},%
\{1,\=1,0,1,0,\=1,0,0\},\{1,\=1,1,\=1,0,1,\=1,0\},\{1,0,\=1,0,0,0,1,1\},\\$~$
\{\=1,0,0,1,0,\=1,0,0\},\{\=1,0,1,\=1,0,1,\=1,0\},%
\{\=1,1,\=1,0,0,0,1,1\},\{0,1,0,0,\=1,0,0,0\},\{1,\=1,1,\=1,1,\=1,0,0\},\\$~$
\{1,0,\=1,0,0,1,\=1,1\},\{1,0,0,0,0,0,1,\=1\},%
\{\=1,0,1,\=1,1,\=1,0,0\},\{\=1,1,\=1,0,0,1,\=1,1\},\{\=1,1,0,0,0,0,1,\=1\},\\$~$
\{0,\=1,0,0,0,0,1,1\},\{1,\=1,1,0,\=1,0,0,0\},%
\{1,0,\=1,0,1,\=1,0,1\},\{1,0,0,0,0,1,\=1,\=1\},\{\=1,0,1,0,\=1,0,0,0\},\\$~$
\{\=1,1,\=1,0,1,\=1,0,1\},\{\=1,1,0,0,0,1,\=1,\=1\},%
\{0,\=1,0,0,0,1,\=1,1\},\{0,\=1,1,0,0,0,1,\=1\},\{1,0,\=1,1,\=1,0,0,1\},\\$~$
\{1,0,0,0,1,\=1,0,\=1\},\{\=1,1,\=1,1,\=1,0,0,1\},%
\{\=1,1,0,0,1,\=1,0,\=1\},\{0,\=1,0,0,1,\=1,0,1\},\{0,\=1,1,0,0,1,\=1,\=1\},\\$~$
\{0,0,\=1,1,0,0,1,0\},\{1,0,0,\=1,0,0,0,1\},%
\{1,0,0,1,\=1,0,0,\=1\},\{\=1,1,0,\=1,0,0,0,1\},\{\=1,1,0,1,\=1,0,0,\=1\},\\$~$
\{0,\=1,0,1,\=1,0,0,1\},\{0,\=1,1,0,1,\=1,0,\=1\},%
\{0,0,\=1,1,0,1,\=1,0\},\{0,0,0,\=1,1,0,1,0\},\{1,0,1,\=1,0,0,0,\=1\},\\$~$
\{\=1,1,1,\=1,0,0,0,\=1\},\{0,\=1,1,\=1,0,0,0,1\},%
\{0,\=1,1,1,\=1,0,0,\=1\},\{0,0,\=1,1,1,\=1,0,0\},\{0,0,0,\=1,1,1,\=1,0\},\\$~$
\{0,0,0,0,\=1,1,1,0\},\{1,1,\=1,0,0,0,0,0\},%
\{\=1,2,\=1,0,0,0,0,0\},\{0,\=1,2,\=1,0,0,0,\=1\},\{0,0,\=1,0,0,0,0,2\},\\$~$
\{0,0,\=1,2,\=1,0,0,0\},\{0,0,0,\=1,2,\=1,0,0\},%
\{0,0,0,0,\=1,2,\=1,0\},\{0,0,0,0,0,\=1,2,0\},\{2,\=1,0,0,0,0,0,0\}\}}
\label{roots}
\einp
\noindent
Replacing
 $\texttt{Flatten[Table[g[m],\{m,0,28\}],1]}\to\texttt{Do[Print[g[m]],\{m,0,58\}]}$
in the last line of {\it Input~(\ref{c:E8})} prints all the roots, at their actual level and produces the characteristic spindle-shaped listing.

\paragraph{\boldmath$E_7$:}
For $E_7$ and $E_6$ the input codes are similar. For $E_7$, the highest root of the adjoint representation, ${\bf 133}$, is $\texttt{\{1,0,0,0,0,0,0\}}$. Its $(133{-}7)/2=63$ positive root vectors of $E_7$ are found by the following code:
\binp
{\tt
ClearAll[a, e, g];\\
 a = \{\{2, \=1, 0, 0, 0, 0, 0\},\\*\hglue10mm
       \{\=1, 2, \=1, 0, 0, 0, 0\},\\*\hglue10mm
       \{0, \=1, 2, \=1, 0, 0, \=1\},\\*\hglue10mm
       \{0, 0, \=1, 2, \=1, 0, 0\},\\*\hglue10mm
       \{0, 0, 0, \=1, 2, \=1, 0\},\\*\hglue10mm
       \{0, 0, 0, 0, \=1, 2, 0\},\\*\hglue10mm
       \{0, 0, \=1, 0, 0, 0, 2\}\};\\
g[0] = \{\{1, 0, 0, 0, 0, 0, 0\}\};\\
g[1] = Table[
   Flatten[g[0]] - a[[Flatten[Position[Flatten[g[0]], 1]][[p]]]], \\*\hglue14mm
   \{p, Length[Flatten[Position[Flatten[g[0]], 1]]]\}];\\
e[x\_] := e[x] = 
   Union[Flatten[\{Table[
       Table[If[g[x][[j]][[i]] == 1, g[x][[j]] - a[[i]], \\*\hglue12mm
         If[g[x][[j]][[i]] == 2, g[x][[j]] - a[[i]]]], \{i, 7\}], 
             \{j, Length[g[x]]\}],\\*\hglue14mm
       Table[Table[
        If[g[x - 1][[l]][[k]] == 2, g[x - 1][[l]] - 2 a[[k]]], \{k, 7\}], 
        \\*\hglue14mm
         \{l, Length[g[x - 1]]\}]\}, 2]];\\
g[x\_] := g[x] = 
  If[MemberQ[e[x - 1], Null], Delete[e[x - 1], 1], e[x - 1]];\\
  Flatten[Table[g[m], \{m, 0, 16\}], 1]}
\label{c:E7}
\einp

\paragraph{\boldmath$E_6$:}
For $E_6$, the highest root (weight of the adjoint representation), ${\bf 78}$ is $\texttt{\{0,0,0,0,0,1\}}$. Its $(78{-}6)/2=36$ positive root vectors are found as follows:
\binp
{\tt
ClearAll[a, e, g];\\
 a = \{\{2, \=1, 0, 0, 0, 0\},\\*\hglue10mm
       \{\=1, 2, \=1, 0, 0, 0\},\\*\hglue10mm
       \{0, \=1, 2, \=1, 0, \=1\},\\*\hglue10mm
       \{0, 0, \=1, 2, \=1, 0\},\\*\hglue10mm
       \{0, 0, 0, \=1, 2, 0\},\\*\hglue10mm
       \{0, 0, \=1, 0, 0, 2\}\};\\
g[0] = \{\{0, 0, 0, 0, 0, 1\}\};\\
g[1] = Table[
   Flatten[g[0]] - a[[Flatten[Position[Flatten[g[0]], 1]][[p]]]], \\*\hglue14mm
   \{p, Length[Flatten[Position[Flatten[g[0]], 1]]]\}];\\
e[x\_] := e[x] = 
   Union[Flatten[\{Table[
       Table[If[g[x][[j]][[i]] == 1, g[x][[j]] - a[[i]], \\*\hglue12mm
         If[g[x][[j]][[i]] == 2, g[x][[j]] - a[[i]]]], \{i, 6\}], \{j, Length[g[x]]\}], 
      \\*\hglue14mm
       Table[Table[
        If[g[x - 1][[l]][[k]] == 2, g[x - 1][[l]] - 2 a[[k]]], \{k, 6\}], \\*\hglue14mm
        \{l, Length[g[x - 1]]\}]\}, 2]];\\
g[x\_] := g[x] = If[MemberQ[e[x - 1], Null], Delete[e[x - 1], 1], e[x - 1]];\\
Flatten[Table[g[m], \{m, 0, 10\}], 1]}
\label{c:E6}
\einp

For the infinite sequences of Lie algebras $A_n,B_n,C_n,D_n$, we recall the low-dimensional isomorphisms\cite{rWyb}
\begin{equation}
  C_1\approx B_1\approx A_1,\qquad
  C_2\approx B_2,\qquad
  D_2\approx A_1\oplus A_1,\qquad
  D_3\approx A_3.
\end{equation}
For this reason, we provide the \textsl{Mathematica} code below as follows:
 $A_n$~for~$n>1$,
 $B_n$ and $C_n$~for~$n>2$,
 $D_n$~for~$n>3$,
and provide the two remaining (low-$n$) cases explicitly, for illustration purposes:
\begin{equation}
  A_1:~ \underbrace{\big[\,2\,\big]}_{\text{Cartan matrix}},\quad
  \underbrace{\texttt{g[0]\,=\,\{\{2\}\}}}_{\text{positive root}},~~
  \underbrace{\texttt{g[1]\,=\,\{\{0\}\}}}_{\text{zero weight}},~~
  \underbrace{\texttt{g[2]\,=\,\{\{\=2\}\}}}_{\text{negative root}},
\end{equation}
which correspond to the well-known $\{J_+,J_z,J_-\}$ generators of $SU_2$.
\begin{equation}
  \begin{array}{r@{\,}c@{\,}l}
   B_2&:& \begin{bmatrix}
            \,2&\bar2\,\\ \,\bar1&2\,
          \end{bmatrix} \\[6mm]
   \bm\a_1 &=& \texttt{\{2,\=2\}}=\text{``$\to$''}\\
   \bm\a_2 &=& \texttt{\{\=1,2\}}=\text{``$\To$''}\\
  \end{array}
  \qquad
  \begin{array}{r@{\,}l}
    \texttt{g[0]\,=}&\quad~~\texttt{\{\{0,2\}%
                      \piC{-6,-1.5}{\rotatebox{-90}{$\To$}}%
                                             \}}\\[2mm]
    \texttt{g[1]\,=}&\quad~~\texttt{\{\{1,0\}%
                      \piC{-10,-.5}{\rotatebox{-135}{$\To$}}%
                      \piC{-4,-1.5}{\rotatebox{-45}{$\to$}}%
                                             \}}%
    \piC{8,0}{$\left.\rule{0pt}{6ex}\right\}$~positive roots}\\[2mm]
    \texttt{g[2]\,=}&\texttt{\{\{2,\=2\}%
                      \piC{-6,-1}{\rotatebox{-90}{$\to$}}%
                                        ,\{\=1,2\}%
                      \piC{-6,-1}{\rotatebox{-90}{$\To$}}%
                                                  \}}\\[2mm]
    \texttt{g[3]\,=}&\texttt{\{\{0,0\}%
                      \piC{-6,-1}{\rotatebox{-90}{$\to$}}%
                                      ,\{0,0\}%
                      \piC{-6,-1}{\rotatebox{-90}{$\To$}}%
                                              \}}%
    \piC{8,0}{zero weights}\\[2mm]
    \texttt{g[4]\,=}&\texttt{\{\{\=2,2\}%
                      \piC{-4,-1.5}{\rotatebox{-45}{$\To$}}%
                                        ,\{1,\=2\}%
                      \piC{-10,-.5}{\rotatebox{-135}{$\to$}}%
                                                  \}}\\[2mm]
    \texttt{g[5]\,=}&\quad~~\texttt{\{\{\=1,0\}%
                      \piC{-6,-1}{\rotatebox{-90}{$\To$}}%
                                             \}}%
    \piC{8,0}{$\left.\rule{0pt}{6ex}\right\}$~negative roots}\\[2mm]
    \texttt{g[6]\,=}&\quad~~\texttt{\{\{0,\=2\}\}}\\
  \end{array}\mkern120mu
 \label{e:B2}
\end{equation}
The Cartan matrix of $C_2$ is the transpose of that of $B_2$, so that the positive simple roots of $C_2$ are the simply the swapped simple roots of $B_2$, whereby the root system of $C_2$ is identical as shown in\eq{e:B2}.

\paragraph{\boldmath$A_n$:}
For $A_n$, the dimension of the adjoint representation is $n(n{+}2)$ and the number of positive root vectors is $(n(n{+}2)-n)/2=n(n{-}1)/2$. The \textsl{Mathematica} code computing the positive root vectors of $A_n$, for $n=5$ for example, is:
\binp
{\tt
ClearAll[n, d, a, g, e];\\
n = 5; (* n = 2, 3, 4, ... *)\\
d = \{\{2, \=1, 0\},\\*\hglue10mm
      \{0, \=1, 2\},\\*\hglue10mm
      \{\=1, 2, \=1\}\};\\
a = If[n > 1, 
   Flatten[\{\{\{PadLeft[d[[1]], n, 0, n - 3]\}\}, \\*\hglue8mm
   \{Table[ PadLeft[ d[[3]], n, 0, n - i - 2], \{i, n - 2\}]\}, \\*\hglue8mm
       \{\{PadRight[ d[[2]], n, 0, n - 3]\}\}\}, 2], \{2\}];\\
g[0] = \{RotateLeft[PadRight[\{1, 1\}, n, 0], 1]\};\\
g[1] = Table[
   Flatten[g[0]] - a[[Flatten[Position[Flatten[g[0]], 1]][[p]]]], \\*\hglue14mm
   \{p, Length[Flatten[Position[Flatten[g[0]], 1]]]\}];\\
e[x\_] := e[x] = 
   Union[Flatten[\{Table[
       Table[If[g[x][[j]][[i]] == 1, g[x][[j]] - a[[i]], \\*\hglue12mm
         If[g[x][[j]][[i]] == 2, g[x][[j]] - a[[i]]]], \{i, n\}], \{j, Length[g[x]]\}], 
      \\*\hglue14mm
      Table[Table[
        If[g[x - 1][[l]][[k]] == 2, g[x - 1][[l]] - 2 a[[k]]], \{k, n\}], \\*\hglue14mm
         \{l, Length[g[x - 1]]\}]\}, 2]];\\
g[x\_] := g[x] = 
   If[MemberQ[e[x - 1], Null], Delete[e[x - 1], 1], e[x - 1]];\\
Flatten[Table[g[m], \{m, 0, n-1\}], 1]}
\label{c:An}
\einp

\paragraph{\boldmath$B_n$:}
For $B_n$, the dimension of the adjoint representation is $n(2n{+}1)$ and the number of positive root vectors is $(n(2n+1)-n)/2=n^2$. The \textsl{Mathematica} code computing the positive root vectors of $B_n$, for $n=5$ for example, is:
\binp
{\tt
ClearAll[n, d, a, g, e];\\
n = 5; (* n = 3, 4, 5, ... *)\\
d = \{\{2, \=1, 0\},\\*\hglue10mm
       \{\=1, 2, \=2\},
\\*\hglue10mm
       \{0, \=1, 2\},\\*\hglue10mm
       \{\=1, 2, \=1\}\};\\
a = 
 Flatten[\{\{\{PadLeft[d[[1]], n, 0, n - 3]\}\}, \\*\hglue8mm
 \{Table[
     PadLeft[d[[4]], n, 0, n - i - 2], \{i, n - 3\}]\}, \\*\hglue8mm
     \{\{PadRight[
      d[[2]], n, 0, n - 3]\}\}, \{\{PadRight[d[[3]], n, 0, n - 3]\}\}\}, 2];\\
g[0] = \{PadRight[\{0, 1, 0\}, n, 0]\};\\
g[1] = \{Flatten[g[0]] - a[[Flatten[Position[Flatten[g[0]], 1]][[1]]]]\};\\
e[x\_] := e[x] = 
   Union[Flatten[\{Table[
       Table[If[g[x][[j]][[i]] == 1, g[x][[j]] - a[[i]],\\*\hglue12mm
         If[g[x][[j]][[i]] == 2, g[x][[j]] - a[[i]]]], \{i, n\}], \{j, 
        Length[g[x]]\}], 
      \\*\hglue14mm
      Table[Table[
        If[g[x - 1][[l]][[k]] == 2, g[x - 1][[l]] - 2 a[[k]]], \{k, n\}], \\*\hglue14mm
         \{l, Length[g[x - 1]]\}]\}, 2]];\\
g[x\_] := g[x] = 
   If[MemberQ[e[x - 1], Null], Delete[e[x - 1], 1], e[x - 1]];\\
Flatten[Table[g[m], \{m, 0, 2n-2\}], 1]}
\label{c:Bn}
\einp

\paragraph{\boldmath$C_n$:}
Similarly to $B_n$, the dimension of the adjoint representation of $C_n$ is also $n(2n{+}1)$ and the number of positive root vectors is also $(n(2n{+}1)-n)/2=n^2$. The \textsl{Mathematica} code computing the positive root vectors of $C_n$, for $n=5$ for example, is:
\binp
{\tt
ClearAll[n, d, a, g, e];\\
n = 5; (* n = 3, 4, 5, ... *) \\
d = \{\{2, \=1, 0\},\\*\hglue10mm
       \{\=1, 2, \=2\},\\*\hglue10mm
       \{0, \=1, 2\},\\*\hglue10mm
       \{\=1, 2, \=1\}\};\\
a = Transpose[Flatten[\{\{\{PadLeft[d[[1]], n, 0, n - 3]\}\},\\*\hglue10mm
 \{Table[PadLeft[d[[4]], n, 0,n - i - 2], \{i, n - 3\}]\},\\*\hglue10mm
 \{\{PadRight[d[[2]], n, 0, n - 3]\}\}, 
\{\{PadRight[d[[3]], n, 0, n - 3]\}\}\}, 2]];\\
g[0] = \{PadRight[\{2\}, n, 0]\};\\
g[1] = \{Flatten[g[0]] - a[[Flatten[Position[Flatten[g[0]], 2]][[1]]]]\};\\
e[x\_] := e[x] = Union[Flatten[\{Table[Table[If[g[x][[j]][[i]] == 1,
 g[x][[j]] - a[[i]],\\*\hglue12mm
If[g[x][[j]][[i]] == 2, g[x][[j]] - a[[i]]]], \{i, n\}], \{j, Length[g[x]]\}],\\*\hglue14mm
Table[Table[If[g[x - 1][[l]][[k]] == 2, g[x - 1][[l]] - 2 a[[k]]], \{k, n\}],\\*\hglue14mm
\{l, Length[g[x - 1]]\}]\}, 2]];\\
g[x\_] := g[x] = Delete[e[x - 1], 1];\\
Flatten[Table[g[m], \{m, 0, 2n-2\}], 1]}
\label{c:Cn}
\einp

\paragraph{\boldmath$D_n$:}
For $D_n$, the dimension of the adjoint representation is $n(2n{-}1)$ and the number of positive root vectors is $(n(2n{-}1)-n)/2=n(n{-}1)$. The \textsl{Mathematica} code computing the positive root vectors of $D_n$, for $n=5$ for example, is:
\binp
{\tt
ClearAll[n, d, a, g, e];\\
n = 5; (* n = 4, 5, 6, ... *) \\
d = \{\{2, \=1, 0, 0\},\\*\hglue10mm
      \{\=1, 2, \=1, \=1\},\\*\hglue10mm
      \{0, \=1, 2, 0\},\\*\hglue10mm
      \{0, \=1, 0, 2\},\\*\hglue10mm
      \{\=1, 2, \=1, 0\}\};\\
a = Flatten[\{\{\{PadLeft[d[[1]], n, 0, n - 4]\}\},\\*\hglue10mm
\{Table[PadLeft[d[[5]], n, 0, n - i - 3], \{i, n - 4\}]\},\\*\hglue10mm
\{\{PadRight[d[[2]], n, 0, n - 4]\}\}, \{\{PadRight[d[[3]], n, 0, n - 4]\}\},\\*\hglue10mm
\{\{PadRight[d[[4]], n, 0, n - 4]\}\}\}, 2];\\
g[0] = \{PadRight[\{0, 1, 0\}, n, 0]\};\\
g[1] = Table[
   Flatten[g[0]] - a[[Flatten[Position[Flatten[g[0]], 1]][[p]]]], \\*\hglue14mm
   \{p, Length[Flatten[Position[Flatten[g[0]], 1]]]\}];\\
e[x\_] := e[x] = 
   Union[Flatten[\{Table[
       Table[If[g[x][[j]][[i]] == 1, g[x][[j]] - a[[i]], \\*\hglue12mm
         If[g[x][[j]][[i]] == 2, g[x][[j]] - a[[i]]]], \{i, n\}], \{j, Length[g[x]]\}], 
      \\*\hglue14mm
      Table[Table[
        If[g[x - 1][[l]][[k]] == 2, g[x - 1][[l]] - 2 a[[k]]], \{k, n\}], \\*\hglue14mm
         \{l, Length[g[x - 1]]\}]\}, 2]];\\
g[x\_] := g[x] = 
   If[MemberQ[e[x - 1], Null], Delete[e[x - 1], 1], e[x - 1]];\\
Flatten[Table[g[m], \{m, 0, 2n-4\}], 1]}
\label{c:Dn}
\einp
\noindent
As with the $E_8$ code {\it Input~(\ref{c:E8})}, replacing
\begin{equation}
 \texttt{Flatten[Table[g[m],\{m,0,{\itshape m$_{\,max}$}\}],1]}~\to~
  \texttt{Do[Print[g[m]],\{m,0,2{\itshape m$_{\,max}$}+2\}]}
\end{equation}
in the last line of the codes {\it Input~(\ref{c:E7})--(\ref{c:Dn})}, where $\texttt{\slshape m$_{\,max}$}$ is the index limit as shown above, prints all the roots at their actual level, forming the characteristic spindle-shaped listing.

The highest root, the level of positive simple root vectors (\ie, the height of the tower of positive roots) and the dimension of the adjoint representation can be found in Table~8 of\cite{rSsky}, while Table~9 of Ref.\cite{rSsky} gives the positive root systems of a few low-rank simple Lie groups. We leave it to the diligent Reader to adapt the above \textsl{Mathematica} codes for the remaining simple Lie groups, $G_2$ and $F_4$.
\ping

In constructing $T^6/\ZZ_N=(\IR^6/\L)/\ZZ_N$ orbifolds for superstring theory and its $M$-theory extension, the choices of the $\ZZ_N$ shift vectors (representing the embedding in the gauge group) are restricted. For example, in $M$-theory, the shift vectors must satisfy a supersymmetry condition, while in string theory they satisfy an additional modular invariance condition; herein, we impose only the former.

We give an example of $\ZZ_7$ vectors. There are 428 eight-component vectors that may be constructed with the components taking values in the standard range $\{0,{1\over7},{2\over7},{3\over7},{4\over7},{5\over7},{6\over7}\}$. The supersymmetry restriction requires that the components of a $\ZZ_N$ vector add up to an integer\cite{rKKKO}. The following code produces all such ``supersymmetric'' $\ZZ_7$-vectors. We have shown only a sample of the output. Note that in order to find all the possible vectors preserving supersymmetry, we need to consider all permutations of the components of each one of the vectors produced by this code; this is accomplished by applying the \textsl{Mathematica} function \texttt{Permutations[{\itshape list}]} to each $\ZZ_7$-vector produced in {\it Output~(\ref{z7v})}, below.

The code under {\it Input~(\ref{z7v})} proceeds as follows:
\begin{enumerate}\itemsep=-3pt\vspace*{-3mm}
 \item[{\tt a}:] stores a list of standard (fractional) nonzero values for the components of the $\ZZ_7$-vectors $\bm{v}$ in\eq{shift}. For general $\ZZ_N$, replace the values with proper fractions $\frac{k}{N}$, for $k=1,\ldots,N$.
 \item[{\tt b}:] stores, for $2\leq i\leq7$, a list of $i$-tuples of possibly repeated component-values from {\tt a}, sorted and with duplicate $i$-tuples removed. For a Lie group of rank $r$, let $2\leq i\leq(r{-}1)$.
 \item[def.:] The list-function {\tt complete[{\itshape list}]} appends the negative of the total sum of the {\tt\itshape list}-components, reduced mod~1, \ie, it appends a (possibly 0) component that makes the total sum into an integer.
 \item[{\tt c}:] applies the list-function ``{\tt complete[{\itshape list}]}'' throughout the list of $i$-tuples ``{\tt b}'', completing them into $i$-tuples with integral totals.
 \item[{\tt q}:] stores the $i$-tuples from ``{\tt c},'' padded by zeros to form 8-vectors, with sorted components, removed duplicates and sorted as vectors. For a Lie group of rank $r$, replace $\texttt{PadRight[c[[i]], 8]}\to\texttt{PadRight[c[[i]], r]}$.
\end{enumerate}\vspace*{-2mm}
To relax the supersymmetric condition for the total sum of the components of the $\ZZ_N$-vectors $\bm{v}$ to be integral, omit line ``{\tt c},'' and replace $\texttt{c}\to\texttt{b}$ in line ``{\tt q}''; the line defining the list-function {\tt complete[{\itshape list}]} thus becomes unused and may also be omitted.

\binp
{\tt
ClearAll[a, b, c, q]; (* Clear arrays from previous computations *)\\
a = \{1/7,2/7,3/7,4/7,5/7,6/7\};\\
b = Union[Sort/@ Flatten[Table[Tuples[a,i],\{i,2,7\}],1]];\\
complete[list\_] := Append[list, Mod[-Total[list], 1]];\\
c = complete/@ b;\\
q = Sort[Union[Sort /@ Table[PadRight[c[[i]], 8], \{i, 1, Length[c]\}]]];\\
"Total no.\ of Z7 Vectors"\\
Length[q]}

\outp
{\tt
\noindent \{\{0,0,0,0,0,0,${1\over7}$,${6\over7}$\}, \{0,0,0,0,0,0,${2\over7}$,${5\over7}$\}, \{0,0,0,0,0,0,${3\over7}$,${4\over7}$\}, \{0,0,0,0,0,${1\over7}$,${1\over7}$,${5\over7}$\},\\$~$
\{0,0,0,0,0,${1\over7}$,${2\over7}$,${4\over7}$\}, \{0,0,0,0,0,${1\over7}$,${3\over7}$,${3\over7}$\}, \{0,0,0,0,0,${2\over7}$,${2\over7}$,${3\over7}$\}, \{0,0,0,0,0,${2\over7}$,${6\over7}$,${6\over7}$\},\\$~$
\{0,0,0,0,0,${3\over7}$,${5\over7}$,${6\over7}$\}, \{0,0,0,0,0,${4\over7}$,${4\over7}$,${6\over7}$\}, \{0,0,0,0,0,${4\over7}$,${5\over7}$,${5\over7}$\}, ..................
\\[2pt]
$~$.................., \{${3\over7}$,${4\over7}$,${4\over7}$,${4\over7}$,${4\over7}$,${4\over7}$,${6\over7}$,${6\over7}$\}, \{${3\over7}$,${4\over7}$,${4\over7}$,${4\over7}$,${4\over7}$,${5\over7}$,${5\over7}$,${6\over7}$\}, \{${3\over7}$,${4\over7}$,${4\over7}$,${4\over7}$,${5\over7}$,${5\over7}$,${5\over7}$,${5\over7}$\},\\$~$
\{${3\over7}$,${4\over7}$,${5\over7}$,${6\over7}$,${6\over7}$,${6\over7}$,${6\over7}$,${6\over7}$\}, \{${3\over7}$,${5\over7}$,${5\over7}$,${5\over7}$,${6\over7}$,${6\over7}$,${6\over7}$,${6\over7}$\}, \{${4\over7}$,${4\over7}$,${4\over7}$,${4\over7}$,${4\over7}$,${4\over7}$,${5\over7}$,${6\over7}$\}, \{${4\over7}$,${4\over7}$,${4\over7}$,${4\over7}$,${4\over7}$,${5\over7}$,${5\over7}$,${5\over7}$\},\\$~$
\{${4\over7}$,${4\over7}$,${4\over7}$,${6\over7}$,${6\over7}$,${6\over7}$,${6\over7}$,${6\over7}$\}, \{${4\over7}$,${4\over7}$,${5\over7}$,${5\over7}$,${6\over7}$,${6\over7}$,${6\over7}$,${6\over7}$\}, \{${4\over7}$,${5\over7}$,${5\over7}$,${5\over7}$,${5\over7}$,${6\over7}$,${6\over7}$,${6\over7}$\}, \{${5\over7}$,${5\over7}$,${5\over7}$,${5\over7}$,${5\over7}$,${5\over7}$,${6\over7}$,${6\over7}$\}\}\\[2mm]
 "Total No. of  Z7 Vectors"\\
428}
\label{z7v}
\einp

One may use a similar code for generating general $\ZZ_N$ shift vectors in the root lattice for $N\neq7$.
\section{Subgroups of $G$}
\label{sgg}
Our next step is to find all the regular, maximal-rank subgroups of $G$, using\eqs{shift}{e:rkHI}.

Our task is indeed closely related to the well-known problem of finding the regular subalgebras of the Lie algebra of $G$, which is accomplished by using the extended Dynkin diagram technique\cite{rD-LieSub}; see also\cite{rWyb,rSsky,rCahn}. The procedure starts with removing in every possible way one node from the extended Dynkin diagram of the Lie algebra of the original group $G$, producing a collection of Dynkin diagrams of the first list of maximal regular subalgebras. One then iterates this procedure for every Lie algebra from this first list. While this procedure is not perfect, all of the very few required corrections are known by now\cite[p.\,135--143]{rCahn}.

\begin{figure}[ht]
\begin{alignat*}9
\unitlength=.5mm
 \begin{picture}(120,20)
\put(-1,0){\circle{5}}
\put(14,0){\circle{5}}
\put(29,0){\circle{5}}
\put(44,0){\circle{5}}
\put(59,0){\circle{5}}
\put(74,0){\circle{5}}
\put(89,0){\circle{5}}
\put(29,15){\circle{5}}
\put(104,0){\circle{5}}
\put(1,0){\line(1,0){11}}
\put(16,0){\line(1,0){11}}
\put(31,0){\line(1,0){11}}
\put(46,0){\line(1,0){11}}
\put(61,0){\line(1,0){11}}
\put(91,0){\line(1,0){11}}
\put(76,0){\line(1,0){11}}
\put(29,2){\line(0,1){11}}
\put(-2,-7){$\SSS\a_1$}
\put(13,-7){$\SSS\a_2$}
\put(28,-7){$\SSS\a_3$}
\put(43,-7){$\SSS\a_4$}
\put(58,-7){$\SSS\a_5$}
\put(73,-7){$\SSS\a_6$}
\put(88,-7){$\SSS\a_7$}
\put(103,-7){$\SSS\a_+$}
\put(32,13){$\SSS\a_8$}
 \end{picture}
&\longrightarrow
\unitlength=.5mm
 \begin{picture}(120,20)(-15,0)
\put(-4,-2){$\times$}
\put(14,0){\circle{5}}
\put(29,0){\circle{5}}
\put(44,0){\circle{5}}
\put(59,0){\circle{5}}
\put(74,0){\circle{5}}
\put(89,0){\circle{5}}
\put(29,15){\circle{5}}
\put(104,0){\circle{5}}
\put(16,0){\line(1,0){11}}
\put(31,0){\line(1,0){11}}
\put(46,0){\line(1,0){11}}
\put(61,0){\line(1,0){11}}
\put(91,0){\line(1,0){11}}
\put(76,0){\line(1,0){11}}
\put(29,2){\line(0,1){11}}
\put(13,-7){$\SSS\a'_7$}
\put(28,-7){$\SSS\a'_6$}
\put(43,-7){$\SSS\a'_5$}
\put(58,-7){$\SSS\a'_4$}
\put(73,-7){$\SSS\a'_3$}
\put(88,-7){$\SSS\a'_2$}
\put(103,-7){$\SSS\a'_1$}
\put(32,13){$\SSS\a'_8$}
\end{picture} \\[3mm]
\unitlength=.5mm
 \begin{picture}(120,20)(0,0)
\put(122,22){\rotatebox{-145}{$\longrightarrow$}}
\put(14,0){\circle{5}}
\put(29,0){\circle{5}}
\put(29,15){\circle{5}}
\put(44,0){\circle{5}}
\put(59,0){\circle{5}}
\put(74,0){\circle{5}}
\put(89,0){\circle{5}}
\put(89,15){\circle{5}}
\put(104,0){\circle{5}}
\put(16,0){\line(1,0){11}}
\put(31,0){\line(1,0){11}}
\put(29,2){\line(0,1){11}}
\put(46,0){\line(1,0){11}}
\put(61,0){\line(1,0){11}}
\put(76,0){\line(1,0){11}}
\put(91,0){\line(1,0){11}}
\put(89,2){\line(0,1){11}}
\put(13,-7){$\SSS\a_7$}
\put(28,-7){$\SSS\a_6$}
\put(43,-7){$\SSS\a_5$}
\put(58,-7){$\SSS\a_4$}
\put(73,-7){$\SSS\a_3$}
\put(88,-7){$\SSS\a_2$}
\put(103,-7){$\SSS\a_1$}
\put(32,13){$\SSS\a_8$}
\put(92,13){$\SSS\a_+$}
 \end{picture}
&\longrightarrow
\unitlength=.5mm
 \begin{picture}(120,20)(-15,0)
\put(14,0){\circle{5}}
\put(29,0){\circle{5}}
\put(29,15){\circle{5}}
\put(44,0){\circle{5}}
\put(56,-2){$\times$}
\put(74,0){\circle{5}}
\put(89,0){\circle{5}}
\put(89,15){\circle{5}}
\put(104,0){\circle{5}}
\put(16,0){\line(1,0){11}}
\put(31,0){\line(1,0){11}}
\put(29,2){\line(0,1){11}}
\put(76,0){\line(1,0){11}}
\put(91,0){\line(1,0){11}}
\put(89,2){\line(0,1){11}}
\put(13,-7){$\SSS\a'_1$}
\put(28,-7){$\SSS\a'_2$}
\put(43,-7){$\SSS\a'_3$}
\put(73,-7){$\SSS\a''_3$}
\put(88,-7){$\SSS\a''_2$}
\put(103,-7){$\SSS\a''_1$}
\put(32,13){$\SSS\a'_4$}
\put(92,13){$\SSS\a''_4$}
 \end{picture}
\end{alignat*}
\caption[Dynkin diagram technique to find $2D_4\subset E_8$]{Removing the node $\a_1$ from the extended Dynkin diagram of $E_8$ (top left) gives the maximal regular subalgebra $D_8$ (top right). Removing the node $\a_4$ from the extended Dynkin diagram of $D_8$ (bottom left) gives the regular subalgebra $2D_4\subset D_8\subset E_8$ (bottom right).}
\label{e8d4d4}
\end{figure}
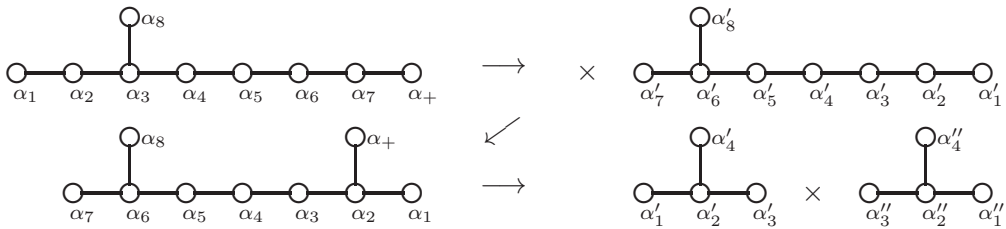

Many of the subalgebras are also found by the quicker method of removing from the extended Dynkin diagram of the group $G$ several nodes in all possible ways at once, and reading off the subalgebra represented by the remainder.
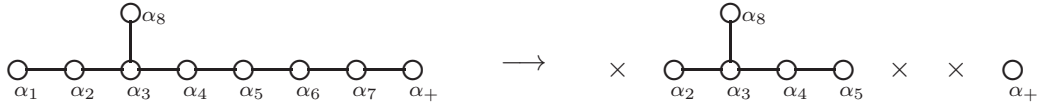
\begin{figure}[ht]
\unitlength=.5mm
$$
 \begin{picture}(120,20)
\put(-1,0){\circle{5}}
\put(14,0){\circle{5}}
\put(29,0){\circle{5}}
\put(44,0){\circle{5}}
\put(59,0){\circle{5}}
\put(74,0){\circle{5}}
\put(89,0){\circle{5}}
\put(29,15){\circle{5}}
\put(104,0){\circle{5}}
\put(1,0){\line(1,0){11}}
\put(16,0){\line(1,0){11}}
\put(31,0){\line(1,0){11}}
\put(46,0){\line(1,0){11}}
\put(61,0){\line(1,0){11}}
\put(91,0){\line(1,0){11}}
\put(76,0){\line(1,0){11}}
\put(29,2){\line(0,1){11}}
\put(-2,-7){$\SSS\a_1$}
\put(13,-7){$\SSS\a_2$}
\put(28,-7){$\SSS\a_3$}
\put(43,-7){$\SSS\a_4$}
\put(58,-7){$\SSS\a_5$}
\put(73,-7){$\SSS\a_6$}
\put(88,-7){$\SSS\a_7$}
\put(103,-7){$\SSS\a_+$}
\put(32,13){$\SSS\a_8$}
 \end{picture}
 ~~\longrightarrow\quad
 \begin{picture}(120,20)
\put(6,-2){$\times$}
\put(24,0){\circle{5}}
\put(39,0){\circle{5}}
\put(54,0){\circle{5}}
\put(69,0){\circle{5}}
\put(39,15){\circle{5}}
\put(81,-2){$\times$}
\put(96,-2){$\times$}
\put(114,0){\circle{5}}
\put(26,0){\line(1,0){11}}
\put(41,0){\line(1,0){11}}
\put(56,0){\line(1,0){11}}
\put(39,2){\line(0,1){11}}
\put(23,-7){$\SSS\a_2$}
\put(38,-7){$\SSS\a_3$}
\put(53,-7){$\SSS\a_4$}
\put(68,-7){$\SSS\a_5$}
\put(42,13){$\SSS\a_8$}
\put(113,-7){$\SSS\a_+$}
 \end{picture}
$$
\caption[Dynkin diagram technique to find regular subgroups of $E_8$]{Removing the nodes $\a_1$, $\a_6$ and $\a_7$ from the extended Dynkin diagram of $E_8$ gives the regular subalgebra $D_5+A_1$. ``$\a_+$'' denotes the extending node; ``$\times$'' denote the locations of the removed nodes.}
\label{e8d5a1}
\end{figure}
For example (see Figure~\ref{e8d5a1}), if we take out the nodes $\a_1$, $\a_6$ and $\a_7$ from the extended Dynkin diagram of $E_8$, we get $D_5+A_1$; see Figure~\ref{e8d5a1}. Notably, however, this does not produce all subalgebras, such as for example $D_4+D_4\subset E_8$, which {\em\/is\/} obtained by the above-outlined iterative method, as shown in Figure~\ref{e8d4d4}.
The resulting complete list of regular subalgebras of $E_8$ has been known since Ref.\cite{rD-LieSub}.

We then pass to the corresponding compact Lie (sub)groups. Rather importantly, the $\ZZ_N$-invariant regular subgroups are necessarily of maximal rank and include $\rk(G)-\rk(\Tw{H}_I)$ abelian factors $U_1$, where $\Tw{H}_I$ is the semisimple factor of the $\ZZ_N$-invariant regular subgroup $H_I\subset G$. This fact renders the centralizer of the semisimple factor in each $\ZZ_N$-invariant subgroup equal to its center, and prevents a direct distinction between inequivalent embeddings of a subgroup; see Appendix~\ref{s:A} for details and a more precise and complete statement. The resulting list of maximal-rank regular subgroups of $E_8$ is given in Table~\ref{atsg}.
\begin{table}[htp]
\caption{Regular subgroups of $E_8$ and their identifiers described in the text. Double daggers ($^\ddag$) indicates subgroups that have two inequivalent embeddings in $E_8$\protect\cite{rD-LieSub}. These are not distinguished by the identification of $\ZZ_N$-invariant subgroups described herein; see Appendix~\ref{s:A} for details.}
\vspace*{-5mm}
$$
\begin{array}{|r|c|c|c|c|c|}
\multicolumn{1}{c}{  } &
\multicolumn{1}{c}{\textbf{Subgroups}} &
\multicolumn{1}{c}{\textbf{p}} &
\multicolumn{1}{c}{\textbf{m}} &
\multicolumn{1}{c}{\textbf{r}} &
\multicolumn{1}{c}{\textbf{c}} \\
\hline
0&E_8&120&0&8&\\\hline\hline\hline
1&SO_{16}&56&0&8&\\\hline
2&SU_9&36&0&8&\Chekk\\\hline
3&SU_8\times SU_2&29&1&8&\\\hline
4&SU_6\times SU_3\times SU_2&19&1&8&\\\hline
5&SU_5^{~2}&20&0&8&\\\hline
6&SO_{10}\times SU_4&26&0&8&\\\hline
7&E_6\times SU_3&39&0&8&\\\hline
8&E_7\times SU_2&64&1&8&\\\hline\noalign{\vglue1mm}\hline
9&SO_{12}\times SU_2^{~2}&32&2&8&\\\hline
10&SO_8\times SU_2^{~4}&16&4&8&\Chekk\\\hline
11&SU_2^{~8}&8&8&8&\\\hline
12&SU_4^{~2}\times SU_2^{~2}&14&2&8&\Chekk\\\hline
13&SO_8^{~2}&24 &0 &8 &\\\hline
14&SU_3^{~4}&12&0&8&\\\hline\hline\hline
15&E_7\times U_1&63&0&7&\Chekk\\\hline
16&SO_{14}\times U_1&42&0&7&\Chekk\\\hline
17&E_6\times SU_2\times U_1&37&1&7&\Chekk\\\hline
18&SO_{12}\times SU_2\times U_1&31&1&7&\\\hline
19&[SU_8]^\ddag\times U_1&28&0&7&\Chekk\\\hline
20&SO_{10}\times SU_3\times U_1&23&0&7&\Chekk\\\hline
21&SO_{10}\times SU_2^{~2}\times U_1&22&2&7&\Chekk\\\hline
22&SU_7\times SU_2\times U_1&22&1&7&\Chekk\\\hline
23&SO_8\times SU_4\times U_1&18&0&7&\\\hline
24&SU_6\times SU_3\times U_1&18&0&7&\\\hline
25&SU_6\times SU_2^{~2}\times U_1&17&2&7&\\\hline
26&SU_5\times SU_4\times U_1&16&0&7&\Chekk\\\hline
27&SO_8\times SU_2^{~3}\times U_1&15&3&7&\Chekk\\\hline
28&SU_5\times SU_3\times SU_2\times U_1&14&1&7&\Chekk\\\hline
29&SU_4^{~2}\times SU_2\times U_1&13&1&7&\\\hline
30&SU_4\times SU_3\times SU_2^{~2}\times U_1&11&2&7&\\\hline
31&SU_3^{~3}\times SU_2\times U_1&10&1&7&\\\hline
32&SU_4\times SU_2^{~4}\times U_1&10&4&7&\\\hline
33&SU_2^{~7}\times U_1&7&7&7&\\\hline \hline
34&E_6\times U_1^{~2}&36&0&6&\Chekk\\\hline
35&SO_{12}\times U_1^{~2}&30&0&6&\Chekk\\\hline
\end{array}
\qquad
\begin{array}{|c|c|c|c|c|c|}
\multicolumn{1}{c}{  } &
\multicolumn{1}{c}{\textbf{Subgroups}} &
\multicolumn{1}{c}{\textbf{p}} &
\multicolumn{1}{c}{\textbf{m}} &
\multicolumn{1}{c}{\textbf{r}} &
\multicolumn{1}{c}{\textbf{c}} \\ \hline
36&SU_7\times U_1^{~2}&21&0&6&\Chekk\\\hline
37&SO_{10}\times SU_2\times U_1^{~2}&21&1&6&\Chekk\\\hline
38&[SU_6]^\ddag\times SU_2\times U_1^{~2}&16&1&6&\Chekk\\\hline
39&SO_8\times SU_3\times U_1^{~2}&15&0&6&\Chekk\\\hline
40&SO_8\times SU_2^{~2}\times U_1^{~2}&14&2&6&\Chekk\\\hline
41&SU_5\times SU_3\times U_1^{~2}&13&0&6&\\\hline
42&[SU_4^{~2}]^\ddag\times U_1^{~2}&12&0&6&\\\hline
43&SU_5\times SU_2^{~2}\times U_1^{~2}&12&2&6&\\\hline
44&SU_4\times SU_3\times SU_2\times U_1^{~2}&10&1&6&\\\hline
45&SU_3^{~3}\times U_1^{~2}&9&0&6&\\\hline
46&SU_4\times SU_2^{~3}\times U_1^{~2}&9&3&6&\\\hline
47&SU_3^{~2}\times SU_2^{~2}\times U_1^{~2}&8&2&6&\\\hline
48&SU_3\times SU_2^{~4}\times U_1^{~2}&7&4&6&\\\hline
49&SU_2^{~6}\times U_1^{~2}&6&6&6&\\\hline \hline
50&SO_{10}\times U_1^{~3}&20&0&5&\\\hline
51&SU_6\times U_1^{~3}&15&0&5&\Chekk\\\hline
52&SO_8\times SU_2\times U_1^{~3}&13&1&5&\\\hline
53&SU_5\times SU_2\times U_1^{~3}&11&1&5&\\\hline
54&SU_4\times SU_3\times U_1^{~3}&9&0&5&\\\hline
55&[SU_4\times SU_2^{~2}]^\ddag\times U_1^{~3}&8&2&5&\\\hline
56&SU_3^{~2}\times SU_2\times U_1^{~3}&7&1&5&\\\hline
57&SU_3\times SU_2^{~3}\times U_1^{~3}&6&3&5&\\\hline
58&SU_2^{~5}\times U_1^{~3}&5&5&5&\\\hline\hline
59&SO_8\times U_1^{~4}&12&0&4&\\\hline
60&SU_5\times U_1^{~4}&10&0&4&\\\hline
61&SU_4\times SU_2\times U_1^{~4}&7&1&4&\\\hline
62&SU_3^{~2}\times U_1^{~4}&6&0&4&\\\hline
63&SU_3\times SU_2^{~2}\times U_1^{~4}&5&2&4&\\\hline
64&[SU_2^{~4}]^\ddag\times U_1^{~4}&4&4&4&\\\hline\hline
65&SU_4\times U_1^{~5}&6&0&3&\\\hline
66&SU_3\times SU_2\times U_1^{~5}&4&1&3&\\\hline
67&SU_2^{~3}\times U_1^{~5}&3&3&3&\\\hline\hline
68&SU_3\times U_1^{~6}&3&0&2&\\\hline
69&SU_2^{~2}\times U_1^{~6}&2&2&2&\\\hline\hline
70&SU_2\times U_1^{~7}&1&1&1&\\\hline\hline
71&U_1^{~8}&0&0&0&\\\hline
\end{array}
$$
\label{atsg}
\end{table}%

For this list of all maximal-rank regular subgroups of $E_8$, we calculate the number of positive root vectors for each subgroup and list them in column {\bf p} of Table~\ref{atsg}. The values of the other identifiers ($r$, $m$ and possibly $m_2,m_3,\dots$) turned out not to be necessary in most cases for our purposes\ft{The identifiers $r$ and $m$ are shown in Table~\ref{atsg} for completeness, and for the benefit of possible generalizations to $\ZZ_N$-actions where the supersymmetry condition is relaxed. The additional identifiers, $m_i$ in {\bfseries Step\,4}, are easily added.}: Before using them, we found the possible candidates which are $\ZZ_7$-invariant subgroups of $E_8$ through a procedure given in {\em\/Input/Output~(\ref{z7roots})\/}.
 This greatly reduced the complexity of the codes in the next section and saves in the \textsl{Mathematica} evaluation time.

	We have 428 $\ZZ_7$ shift vectors in {\em Output (\ref{z7v})} and once we take their permutations, this gives a total of 823,542 shift vectors. We take the first $\ZZ_7$ vector {\tt\{0,0,0,0,0,0,${1\over7}$,${6\over7}$\}} from the previous section and calculate the number of positive root vectors that satisfy the condition $\bm{p}{\cdot}\bm{v}\in\ZZ$ using the following code:
\binp
{\tt
q = (not shown here: 428 $\mathbb {Z}_7$ vectors from Output (\ref{z7v}));\\
p = (not shown here: 120 positive roots of $E_8$ from Output(\ref{roots}));\\
v = Flatten[Table[Permutations[q[[i]]], \{i, 1, 1\}], 1];\\
u = Table[Table[p[[i]].v[[j]], \{i, Length[p]\}], \{j, Length[v]\}];\\
w = Table[
   Table[IntegerQ[u[[j, i]]], \{i, Length[p]\}], \{j, Length[v]\}];\\
r = Table[Count[w[[j]], True], \{j, Length[v]\}];\\
Union[r] >>> Z7\_Roots;}

\outp
\noindent{\tt \{37,42\}}
\label{z7roots}
\einp

The code under {\it Input~(\ref{z7roots})} reads
 the 428 $\ZZ_7$-vectors from {\it Output~(\ref{z7v})} into the list ``{\tt q}'' and
 the 120 positive roots from {\it Output~(\ref{roots})} into the list ``{\tt p}'' and then proceeds as follows:\\
{\tt v}: The list of $\ZZ_7$-vectors $\bm{v}$ obtained as permutations of the first vector in {\it Output~(\ref{z7v})}.\\
{\tt u}: Stores the dot products between each one of the vectors from ``{\tt v}'' and the 120 positive roots in ``{\tt p}''.\\
{\tt w}: Finds the integral dot products in ``{\tt u}''.\\
{\tt r}: Counts the number of integral dot products in ``{\tt u}'', which is the total number of positive roots (in ``{\tt p}'') satisfying the condition {\boldmath $p.v$} $\in \mathbb{Z}$, for each one of the vectors {\boldmath$v$} in ``{\tt v}''.

The {\em Output} of this evaluation ({\tt \{37,42\}}) is written in an external file~\Fl{Z7\_Roots}. We do this evaluation for the other $\ZZ_7$ vectors in {\em Output (\ref{z7v})} and the results are collected from the text file~\Fl{Z7\_Roots}. This gives the possible values of the identifier $p$ for $\ZZ_7$ vectors as
\begin{equation}
{\tt p: \{14, 15, 16, 21, 22, 23, 28, 30, 36, 37, 42, 63\}}
\label{valp}
\end{equation}
This narrows down our choices to 21 subgroups of $E_8$, marked by a check in column {\bf c} of Table \ref{atsg}. Now we use the values of $m$ (number of  $A_1$ factors in $H_I\subset E_8$) to identify the possible subgroups $H_I\subset E_8$. When $p$ and $m$ do not specify $H_I\subset E_8$ unambiguously, we use the values of $r$ (rank of the semisimple part of $H_I$). The values of the identifiers $p$, $m$ and $r$ are also calculated from the root vectors that are invariant\eq{shift} with respect to a $\ZZ_7$ shift. This is shown in the next section.
\section{$\ZZ_N$ Invariant Subgroups of $G$}
\label{znisg}
To illustrate the procedure of calculating the values of $m$ and $r$ from the $\ZZ_7$-invariant root vectors we give the same example as in Section 3 of our previous paper\cite{rAHZ7}. Take the shift vector {\tt $\bm{v}=\{{1\over7},{1\over7},0,{2\over7},0,0,{3\over7},0\}$}, which is one of the permutations of {\tt \{0,0,0,0,${1\over7}$,${1\over7}$,${2\over7}$,${3\over7}$\}}. The $\ZZ_7$-invariant $E_8$ root vectors are
\begin{equation}
\begin{gathered}
\texttt{\{0,0,0,0,1,\=1,0,0\},~\{1,\=1,0,0,0,0,0,1\},~\{1,\=1,1,0,0,0,0,\=1\}},\\
\texttt{\{\=1,1,0,0,\=1,1,0,0\},~\{0,0,\=1,0,1,0,0,1\},~\{0,0,0,0,1,0,0,\=1\}},\\
\texttt{\{1,\=1,0,0,0,1,0,0\},~\{0,0,0,0,0,\=1,0,1\},~\{0,0,1,0,0,\=1,0,\=1\}},\\
\texttt{\{0,1,\=1,1,\=1,1,\=1,0\},~\{\=1,0,1,\=1,0,0,1,0\},~\{1,\=1,1,0,\=1,0,0,0\}},\\
\texttt{\{\=1,1,\=1,0,1,\=1,0,1\},~\{\=1,1,0,0,1,\=1,0,\=1\},~\{0,0,\=1,0,0,0,0,2\}}
\end{gathered}
\end{equation}
and are thus invariant under the action of the group $\ZZ_7$ generated by this shift.
Call these root vectors $t[i]$, $i=1,2,\cdots15$, and set $p=15$.

Next, we need to identify which subgroup $H_I\subset E_8$|from among those listed in Table~\ref{atsg}|do these roots (together with their negatives and the Cartan root vectors) generate.
We look for possible relations in the form $t[i]+t[j]=t[k]$ and find the following:
\begin{subequations}
\begin{alignat}9
 t[2]+t[14]&=t[1],&\qquad
 t[3]+t[13]&=t[1],&\qquad
 t[5]+t[9] &=t[1],&\qquad
 t[6]+t[8] &=t[1],
 \label{e:D4L5}\\[2mm]
 t[3]+t[15]&=t[2],&\qquad
 t[5]+t[12]&=t[2],&\qquad
 t[7]+t[8]&=t[2],
 \label{e:D4L4}\\[2mm]
 \begin{array}{r@{}}
  t[6]+t[12]\\ t[7]+t[9]\\
 \end{array}
 &\begin{array}{l@{}}
  =t[3]\\ =t[3]\\
 \end{array}\bigg\}&\qquad
 \begin{array}{r@{}}
  t[6]+t[15]\\ t[7]+t[13]\\
 \end{array}
 &\begin{array}{l@{}}
  =t[5]\\ =t[5]\\
 \end{array}\bigg\}&\qquad
 \begin{array}{r@{}}
  t[9]+t[15]\\ t[12]+t[13]\\
 \end{array}
 &\begin{array}{l@{}}
  =t[8]\\ =t[8]\\
 \end{array}\bigg\}
 \label{e:D4L3}\\[2mm]
 t[7]+t[14]&=t[6],&\qquad
 t[12]+t[14] &=t[9],&\qquad
 t[14]+t[15] &=t[13],
 \label{e:D4L2}\\[2mm]
 t[10]+t[11]&=t[4].
 \label{e:A2}
\end{alignat}
\label{rrln}
\end{subequations}
Since the root vectors $t[7],\,t[10],\,t[11],\,t[12],\,t[14]$ and $t[15]$ cannot be expressed as a sum of any other root vectors, they must correspond to 6 positive, simple root vectors in $H_I$. The rank of the semisimple part of $H_I$ then must be 6, and the remaining two zero weights correspond to a $U(1)^2$ factor. Also, all the 15 root vectors appear in (\ref{rrln}), meaning this $H_I$ has no $A_1$ factors, each of which would have had to have a single, isolated, positive root vector. From these $\ZZ_7$-invariant root vectors the variables $m$ and $r$ are defined as:
\begin{quote}
\begin{description}\vspace*{-2mm}
 \item[$m$] is the number of root vectors that do not appear in the equation of the form $t[i]+t[j]=t[k]$ and so must be single, isolated, positive root vectors; here, $m=0$.
 \item[$r$] is the number of root vectors that do not appear on the right side of the relations of the form $t[i]+t[j]=t[k]$ and so must be simple; here, $r=6$.
\end{description}\vspace*{-2mm}
\end{quote}

Using $\{p,m,r\}=\{15,0,6\}$, we identify unambiguously the subgroup from Table~\ref{atsg} as $SO_8\times SU_3$.

Observe that this is indeed consistent with the structure of the relations\eq{rrln}:
\begin{enumerate}\vspace*{-2mm}
 \item The positive roots $t[4]$, $t[10]$ and $t[11]$ form a separate rank-2 positive root system where $t[10]$ and $t[11]$ are simple and $t[4]$ is their sum\eq{e:A2}; this can correspond only to $SU_3$.
 \item The positive roots $t[6]$, $t[9]$ and $t[13]$ are each obtained as a sum\eq{e:D4L2} of two of the positive simple roots $\{t[7],t[12],t[14],t[15]\}$, and so must be one level above these positive simple roots.
 \item Expressing $t[6]$, $t[9]$ and $t[13]$ in this way,
  $t[3]$, $t[5]$ and $t[8]$ are each found to be a sum\eq{e:D4L3} of three of the positive simple roots, and so are two levels above the positive simple roots.
 \item In this way, $t[2]=t[7]+t[12]+t[14]+t[15]$ is a sum\eq{e:D4L4} of all four distinct positive simple roots,
  while $t[1]=t[2]+t[14]=t[7]+t[12]+2\,t[14]+t[15]$ has one more positive simple root\eq{e:D4L5}. Therefore, $t[2]$ and $t[1]$ occupy respectively the third and fourth level above the positive simple roots.
\end{enumerate}\vspace*{-2mm}
These facts are consistent with $\{t[7],t[12],t[14],t[15];\,t[6],t[9],t[13];\,t[3],t[5],t[8];\,t[2];\,t[1]\}$ forming the po\-si\-tive root system of $SO(8)$, \ie, its Lie algebra $D_4$. As it turns out, such a more detailed study was not needed in determining the list of $\ZZ_7$-invariant subgroups of $E_8$ in Table~\ref{giz7s} and the identifiers $\{p,m,r\}$ did suffice to this end.

We employ this analysis in the construction of the \textsl{Mathematica} codes below and using the identifiers $\{p,m,r\}$ identify the fourteen subgroups of $E_8$ that are invariant under a $\ZZ_7$ shift listed in Table~\ref{giz7s}, and so in fact the complete $\ZZ_7$ group action generated by that shift.
\begin{table}[ht]
\caption{$\ZZ_7$-invariant subroups of $E_8$.} 
\label{giz7s}
\begin{center}
\begin{tabular}{|c|c|c|c|c|c|c|c|}
\hline
&Group&& Group&&Group&&Group\\\hline
1&$E_7$&5&$SO_{12}$&9&$SU_8$&13&$SU_5\times SU_4$\\\hline
2&$E_6\times SU_2$&6&$SO_{10}\times SU_3$&10&$SU_7\times SU_2$&14&$SU_5\times SU_3\times SU_2$\\\hline
3&$E_6$&7&$SO_{10}\times SU_2$&11&$SU_7$&&\\\hline
4&$SO_{14}$&8&$SO_8\times SU_3$&12&$SU_6\times SU_2$&&\\\hline
\end{tabular}\\[2mm]
\end{center}
\end{table}

\binp
{\tt
q = (not shown here: 428 $\mathbb {Z}_7$ vectors from Output (\ref{z7v}));\\
CleanSlate[]:\\
v = Flatten[Table[Permutations[q[[i]]], \{i, 1, 1\}], 1];\\
u = Table[Table[p[[i]].v[[j]], \{i, Length[p]\}], \{j, Length[v]\}];\\
w = Table[
   Table[IntegerQ[u[[j, i]]], \{i, Length[p]\}], \{j, Length[v]\}];\\
s = Table[Flatten[Position[w[[k]], True]], \{k, Length[w]\}];\\
t = Table[
   Table[p[[s[[j]][[i]]]], \{i, Length[s[[j]]]\}], \{j, Length[w]\}];\\
$\Phi$[k\_] := $\Phi$[k] = 
   Evaluate[
    b = Table[
      Table[t[[k]][[i]] + t[[k]][[j]],\\\hspace*{3cm} \{i, Length[t[[k]]]\}], \{j, 
       Length[t[[k]]]\}];\\
    c = Table[
      Table[MemberQ[t[[k]], b[[i, j]]], \{i, Length[t[[k]]]\}], \{j, 
       Length[t[[k]]]\}];\\
    f = Position[c, True];\\
    g = Union[Table[Sort[f[[i]]], \{i, Length[f]\}]];\\
    x = Table[g[[i]][[1]], \{i, Length[g]\}];\\
    y = Table[g[[i]][[2]], \{i, Length[g]\}];\\
    h = Table[t[[k]][[x[[i]]]] + t[[k]][[y[[i]]]], \{i, Length[x]\}];\\
    z = Flatten[Table[Position[t[[k]], h[[i]]], \{i, Length[h]\}]];\\
    o = Table[l, \{l, Length[t[[k]]]\}];\\
    m = Length[Complement[o, Union[x, y, z]]];\\
    r = Length[Complement[o, z]];];\\
Table[If[Length[t[[k]]] == 14, 
   Evaluate[$\Phi$[k]; 
    If[m == 2, 
     If[r == 6, a[1] a[1] d[4],\\\hspace*{6cm} If[r == 8, a[1] a[1] a[3] a[3]]], 
     a[1] a[2] a[4]]], \\
   If[Length[t[[k]]] == 15, 
    Evaluate[$\Phi$[k]; 
     If[m == 0, If[r == 5, a[5], \\\hspace*{5cm}If[r == 6, a[2] d[4]]], 
      a[1] a[1] a[1] d[4]]], \\
    If[Length[t[[k]]] == 16, 
     Evaluate[$\Phi$[k]; 
      If[m == 0, a[3] a[4], 
       If[m == 1, a[1] a[5], \\\hspace*{5cm} If[m == 4, a[1] a[1] a[1] a[1] d[4]]]]], \\
     If[Length[t[[k]]] == 21, 
      Evaluate[$\Phi$[k]; If[m == 0, a[6], a[1] d[5]]], \\
      If[Length[t[[k]]] == 22, 
       Evaluate[$\Phi$[k]; 
        If[m == 1, a[1] a[6], a[1] a[1] d[5]]], \\
       If[Length[t[[k]]] == 23, a[2] d[5], \\
        If[Length[t[[k]]] == 28, a[7], \\
         If[Length[t[[k]]] == 30, d[6], \\
          If[Length[t[[k]]] == 36, 
           Evaluate[$\Phi$[k]; If[r == 6, e[6], a[8]]], \\
           If[Length[t[[k]]] == 37, a[1] e[6], \\
            If[Length[t[[k]]] == 42, d[7], \\
             If[Length[t[[k]]] == 63, e[7]]]]]]]]]]]]], \{k, 
   Length[t]\}];\\
Union[\%]>>>Z7\_Groups;}

\outp
{\tt
\noindent \{d[7], a[1]\*e[6]\}}
\label{evalt}
\einp

The code under {\it Input~(\ref{evalt})} reads
 the 428 $\ZZ_7$-vectors from {\it Output~(\ref{z7v})} into the list ``{\tt q}'' and
 the 120 positive roots from {\it Output~(\ref{roots})} into the list ``{\tt p}'' and then proceeds as follows:\\
{\tt v, u, w}: Have the same meaning as in Input (9).\\
{\tt s, t}: For each one of the $\mathbb{Z}_7$ shift-vectors in ``{\tt v}'', these find the set of positive root vectors of $E_8$ that have integral scalar products with the $\mathbb{Z}_7$ shift-vector.\\
{\tt $\Phi$}: This function uses the analysis as stated after equation (5.1) to find the values of $m$ and $r$ as defined in section 2. Note that {\tt $\Phi$} is a function which uses the variables ``{\tt c}'', ``{\tt f}'', ``{\tt g}'', ``{\tt x}'', ``{\tt y}'', ``{\tt h}'', ``{\tt z}'' and ``{\tt o}'' to evaluate ``{\tt m}'' and ``{\tt r}'' (which have the meaning of $m$ and $r$ from Section~\ref{algo}). This function is evaluated only when the number of $\ZZ_7$-invariant roots of $E_8$ (in the code this number is {\tt Length[t[[k]]]}) is not enough to identify the subgroup $H_I$ as discussed in this section.
 The quantity {\tt Length[t[[k]]]} is the number of $\ZZ_7$-invariant root vectors ($=p$), ``{\tt m}'' is the number of $A_1$ factors ($=m$) and ``{\tt r}'' is the rank ($r$) of a group. These three variables are calculated from the $\ZZ_7$-invariant root vectors as explained in the above example, the $E_8\supset SO(8)\times SU(3)$ subgroup. The output of this evaluation is written in an external file~\Fl{Z7\_Groups} where a group $A_n$ is identified as {\tt a[n]}, $D_n$ as {\tt d[n]} and $E_n$ as {\tt e[n]}.

For other orbifolds there are situations where $p$, $m$ and $r$ do not suffice to specify the group unambiguously. In those cases we look for $A_2$, $A_3$, $\cdots$ factors in $H_I$ by looking at root vector relations. For example, an $A_2$ factor would have to be spanned by three root vectors $\{t[i],t[j],t[k]\}$ that satisfy an equation of the form $t[i]+t[j]=t[k]$ and occur in no equation involving any other root vectors. Equivalently, we can look for root vectors that do not appear in any equation of the form $t[i]+t[j]+t[k]=t[l]$.
\section{Automation}
\label{autom}
Due to limitations of computer's processor speed and memory, it may be necessary to partition the computation. The following shows how it may be done for the $\ZZ_7$ orbifold example in $M$-theory.

\paragraph{\bfseries(i)} Collect all the $\ZZ_7$ vectors {\tt q} in {\em Output (\ref{z7v})} and all the $E_8$ root vectors {\tt p} in {\em Output (\ref{roots})} of section \ref{rsv} and put them in a notebook, say, \Fl{NB\_0}. Use the package `CleanSlate'\ft{This package is available on-line at: {\tt http://library.wolfram.com/infocenter/MathSource/4718/}.} and put this in one of \textsl{Mathematica}'s home directory (\texttt{\$HomeDirectory}). This package helps in clearing the \textsl{Mathematica} kernel memory so that successive evaluations can use the maximum possible memory. The input of~\Fl{NB\_0} are as follows:

\binp
{\tt
\noindent q = ; \textrm{({\em no output shown here: 428 $\ZZ_7$ vectors from Output (\ref{z7v})}~)}\\
 p = ; \textrm{({\em no output shown here: 120 positive root vectors of $E_8$ from Output (\ref{roots})}~)}\\
 << CleanSlate.m;\\
  orbifold = EvaluationNotebook[];\\
 NotebookSave[orbifold]\\
 NotebookOpen["NB\_1.nb"]}
 \label{nb0}
 \einp
 
\paragraph{\bfseries(ii)} We create a notebook~\Fl{Z7\_Generic} in \texttt{\$HomeDirectory} which contains the code of {\em Input (\ref{z7roots})} with some added lines of codes to make use of the automation process:

\binp
{\tt
NotebookClose[orbifold]\\
CleanSlate[];\\
 v = Flatten[Table[Permutations[q[[i]]], \{i, $\a$, $\a$\}], 1];\\
 u = Table[Table[p[[i]].v[[j]], \{i, Length[p]\}], \{j, Length[v]\}];\\
 w = Table[ Table[IntegerQ[u[[j, i]]], \{i, Length[p]\}], \{j, Length[v]\}];\\
 r = Table[Count[w[[j]], True], \{j, Length[v]\}];\\
Union[r] >>> Z7\_Roots;\\
orbifold = EvaluationNotebook[];\\
NotebookSave[orbifold]\\
$\g$ = $\a$ + 1;\\
"NB " <> ToString[$\g$\,] <> ".nb";\\
InputForm[\%]\\
NotebookOpen[\%];}
 \label{gener}
 \einp
 
 \noindent Next we create a notebook~\Fl{Z7\_Generator} with the following set of codes,
 
\binp
{\tt
Do[NotebookPut[NotebookGet[First[Notebooks["Z7\_Generic.nb"]]]/."$\a$"->$\b$];\\
NotebookSave[SelectedNotebook[],"NB "<>ToString[$\b$]<>".nb"];\\
Pause[2];\\
NotebookClose[SelectedNotebook[]],{$\b$,1,428}]}
\label{noteb}
\einp

\noindent Once the {\em Input (\ref{noteb})} is run, it creates 428 notebooks with the contents of {\em Input (\ref{gener})} where the value of $\a=1,2,3,\cdots,428$, respectively, for each notebook. The files are created in the {\tt \$HomeDirectory}.

\paragraph{\bfseries(iii)} Our next step is to evaluate these 428 notebooks in a way such that when we open~\Fl{NB\_0}, it automatically evaluates it's content and the contents of notebooks~\Fl{NB\_1}, \Fl{NB\_2} and so on so forth. The {\tt NotebookClose[orbifold]} input line closes the previous notebook that has been evaluated. In this way the screen is not cluttered with open \textsl{Mathematica} notebooks, improving the performance of the computer's memory. The memory is also  managed by the input line {\tt CleanSlate[]}. Note that the `CleanSlate' package is called in after \textsl{Mathematica} stores the values of {\tt q} and {\tt w} in its memory which is necessary for the whole evaluation process. The end result is  collected from the text file~\Fl{Z7\_Roots} created in {\tt \$HomeDirectory} and is given in \Eq{valp}.

\paragraph{\bfseries(iv)} We apply a similar procedure for the evaluation of the $\ZZ_7$ invariant groups, {\em Input (\ref{evalt})}. The results are collected from the text file~\Fl{Z7\_Groups} and are summerized in Table \ref{giz7s}.

In order for the automation process to work we need to make the following changes to \textsl{Mathematica} preferences,

\noindent 1.~\sl
 Notebook Options $\rightarrow$ File Options $\rightarrow$ Notebook Autosave (False $\rightarrow$ True)\\ \rm
2.~\sl Notebook Options $\rightarrow$ File Options $\rightarrow$ ClosingAutosave (False $\rightarrow$ True)\\ \rm
3.~\sl Notebook Options $\rightarrow$ File Options $\rightarrow$ AutogeneratedPackage (Manual $\rightarrow$ None)\\*[-2mm]
\noindent\hbox to\hsize{\hrulefill}\\ \rm
\noindent 4.~\sl Notebook Options $\rightarrow$ Evaluation Options $\rightarrow$  Initialization CellEvaluation (Automatic $\rightarrow$ True)\\ \rm
5.~\sl Notebook Options $\rightarrow$ Evaluation Options $\rightarrow$ Initialization CellWarning (True $\rightarrow$ False)\\*[-2mm]
\noindent\hbox to\hsize{\hrulefill}\\ \rm
\noindent 6.~\sl Cell Options $\rightarrow$ Evaluation Options $\rightarrow$ Initialization Cell (False $\rightarrow$ True)\\ \rm

This automation process was first tested and used in version 5.2 of \textsl{Mathematica}, where it worked as designed. For later versions, there appears to be a problem which prevents the evaluation of a notebook when it is opened by another notebook, even though the \textsl{Initialization CellEvaluation} and \textsl{Initialization Cell} are changed to \textsl{True} (globally). In those versions of \textsl{Mathematica}, the automation process \textsf{\bfseries(iii)} can be performed using a code such as:

\binp
{\tt
nb = NotebookOpen["notebook.nb"];\\
SelectionMove[nb, All, Notebook];\\
SelectionEvaluate[nb];\\
orbifold = EvaluationNotebook[];\\
NotebookSave[orbifold];\\
NotebookClose[orbifold];}
 \label{autom2}
\einp
\noindent
Corresponding changes need to be made also in {\em Input (\ref{nb0})} and {\em Input (\ref{gener})} for this automation process to work.
\section{Conclusion}
\label{con}
We have shown in detail how to find the $\ZZ_7$-invariant subgroups of $E_8$ using \textsl{Mathematica}.
 These groups, obtained in orbifold $M$-theory, turn out to be closely related to string theory compactification down to four dimensions:
 In the limit $x^{11}\rightarrow 0$, the two $\ZZ_7$-invariant subgroups of $E_8$ (one on each of the two boundaries of $x^{11}$) coalesce into $H_{I,\sss L}\times H_{I',\sss R}$, which turn out to coincide with the gauge groups found in $\ZZ_7$-orbifold models in string theory\cite{rKKKO}.  We have tested our codes also for $\ZZ_2$, $\ZZ_3$, $\ZZ_4$ and $\ZZ_6$ orbifolds. The so-obtained subgroups upon the limit $x^{11}\rightarrow 0$ coincide with those found in string theory compactification. This would imply that our codes can also be used for $\ZZ_8$ and $\ZZ_{12}$ orbifolds.

In the presence of gauge background fields (Wilson lines) the four-dimensional gauge group breaks down to some smaller groups. Since these Wilson lines provide additional shifts in the group lattice, it should be possible to employ our procedure also in those types of models.

For the simple Lie groups $A_n, B_n, C_n, D_n, E_6$ and $E_7$, our procedure can be applied in finding the unbroken gauge symmetry under any $\ZZ_N$ shifts. In section \ref{rsv}, we provided the root vectors for these groups. As semisimple Lie groups are products of simple Lie groups, the procedure merely needs to be applied to each factor separately.

Finally, our present goal was the demonstration that \textsl{Mathematica} can be used to compute $\D$-invariant subgroups of semisimple Lie groups. In achieving this goal, several additional topics came to our attention, which provide ground for further investigation.
 In particular, having been motivated by applications in $M$-theory and also for simplicity, we have restricted our attention to ``supersymmetric'' $\D$-actions and moreover to $\D=\ZZ_N$.
 Secondly, the analysis as presented herein does not distinguish between inequivalent embeddings of a subgroup $H_I$ within the original Lie group.
 Lastly, it would seem desirable to re-structure and package the computations presented herein into a single, interactive \textsl{Mathematica} package.
 Generalizations of our work in each of these directions would seem to be quite worthwhile, but are beyond our present scope and we defer this to a separate effort.

\bigskip
\paragraph{\bfseries Acknowledgment:}
 We should like to thank the Referee, Prof.~Todor Milev,
 for the superb work and constructive criticism in reviewing the initial version of this article, and for
 pointing out a serious error in some of the intermediate results. Although their correction turns out not to change the final result (Table~\ref{giz7s}), it did provide an opportunity not only to present our results correctly but also to better explain the details of the work and to clarify the subtleties in identifying the maximal-rank subgroups; see Appendix~\ref{s:A}.
 We are indebted to the generous support by the Department of Energy through the grant DE-FG02-94ER-40854. T.H.\ wishes to thank for the recurring hospitality and resources provided by
 the Physics Department of the University of Central Florida, Orlando, and
 the Physics Department of the Faculty of Natural Sciences of the University of Novi Sad, Serbia, where part of this work was completed.

\appendix
\section{Regular Subalgebras of $E_8$}
\label{s:A}
Physics applications in grand-unified model building\cite{rSsky} and string-theory and its $M$-theory extension\cite{rPHEW1,rBBS} focus on compact classical Lie groups, and often also on an application-dependently restricted subset of their lowest-dimensional unitary representations. Such is the case in Refs.\cite{rPHEW1,rAHZ7,rKKKO}, where
 ({\small\bf1})~only the adjoint representation of $E_8$ is considered, and
 ({\small\bf2})~only the $\ZZ_N$-invariant subgroups $H_I$.
In particular, the $\ZZ_N$-invariant subgroups $H_I\subset G$ all satisfy\eqs{shift}{e:rkHI} and have their centralizer equal its center; see below. Also, finite factors and the real forms of the Lie groups are not considered and we easily pass from Lie algebras to the corresponding compact Lie groups.

\subsection{Subalgebras}
An exhaustive procedure for listing the regular subalgebras of Lie algebras was provided originally by E.B.~Dynkin\cite{rD-LieSub}, is well described in texts\cite{rJH-ILieA+RT,rWyb,rCahn}, review literature\cite{rSsky} and also in research articles such as Ref.\cite{rKKKO}. One starts with listing the maximal semisimple regular subalgebras by removing one node from the extended Dynkin diagram of the original algebra. For $E_8$, these are\cite{rD-LieSub}:
\begin{equation}
  E_8 ~\supset~
  D_8,~~
  A_8,~~
  A_7+A_1,~~
  A_5+A_2+A_1,~~
  2A_4,~~
  D_5+A_3,~~
  E_6+A_2~~\text{and}~~
  E_7+A_1.
 \label{e:E8mSSrsa}
\end{equation}
Next, proceed by listing the maximal semisimple regular subalgebras of\eq{e:E8mSSrsa}, and continue so iteratively. This adds
\begin{equation}
  D_6+2A_1,~~
  D_4+4A_1,~~
  8A_1,~~
  2A_3+2A_1,~~
  2D_4~~\text{and}~~
  4A_2
\label{e:E8SSrsa}
\end{equation}
to the list\eq{e:E8mSSrsa}, completing the list of all semisimple regular subalgebras of maximal rank\cite[Table~10, p.\,147]{rD-LieSub}.
 Non-semisimple maximal subalgebras are now found by applying to the list\eqs{e:E8mSSrsa}{e:E8SSrsa} the results in Dynkin's Table~12.a\cite[p.\,151]{rD-LieSub}:
\begin{equation}
  \begin{array}{r@{\>\supset\>}l@{\qquad}cr@{\>\supset\>}l@{\qquad}cr@{\>\supset\>}l}
 A_n & A_k+A_{n-k-1}+K_1,          && B_n & B_{n-1}+K_1, && C_n & A_{n-1}+K_1,\\ 
 D_n & D_{n-1}+K_1,~~ A_{n-1}+K_1, && E_6 & D_5+K_1,     && E_7 & E_6+K_1, \\ 
  \end{array}
 \label{e:nSSsa}
\end{equation}
where $k=0,1,2,\ldots,n{-}2$ for $n>1$ and $A_0\define\varnothing$, and $K_1$ is the ``null algebra'' consisting of a single Cartan element, generating an abelian factor $U(1)$ in the corresponding Lie subgroup. For $E_8$, this produces the listing
\begin{equation}
 \begin{aligned}
  \bigg\{
  \begin{array}{ccccccc}
A'_7,  &A_6{+}A_1, &A_5{+}2A_1, &A_4{+}A_3, &A_4{+}A_2{+}A_1, &A_3{+}A_2{+}2A_1,
  &A_3{+}4A_1\\[0mm]
A''_7, &E_6{+}A_1, &D_7, &D_5{+}A_2, &D_5{+}2A_1, &D_4{+}A_3,
  &3A_2{+}A_1\\[0mm]
  \end{array}\bigg\}+K_1,\\
 \big\{A_4{+}2A_1,~~ D_4{+}A_2,~~ 2A_3,~~ 2A_2{+}2A_1,~~ A_2{+}4A_1\big\}+2K_1,
 \end{aligned}
 \label{e:E8nSSrsa}
\end{equation}
omitting the non-semisimple subalgebras wherein a $K_1$ summand is subsumed within a proper $A_1$ summand in an otherwise identical subalgebra in the listing. The two separate copies of $A_7+K_1$ however are listed as inequivalent subalgebras, in that $A'_7{+}K_1\subset A_8\subset E_8$ whereas $A''_7{+}K_1\not\subset A_8\subset E_8$\cite{rD-LieSub}, which is easily traced in the progression from\eq{e:E8mSSrsa} to\eq{e:E8SSrsa} to\eq{e:E8nSSrsa}. 

Finally, in addition to the combined listing of $8+6+19$ subalgebras\eqs{e:E8mSSrsa}{e:E8SSrsa}--(\ref{e:E8nSSrsa}), the remaining 42 subalgebras are obtained by omitting summands from the entries\eqs{e:E8mSSrsa}{e:E8SSrsa}--(\ref{e:E8nSSrsa}) in all possible ways. In doing so, one must take into account that the omitted summands may turn out to be subsumed in the (larger) centralizer in $E_8$, inducing an equivalence of the remaining summand(s). For example, already in the list\eq{e:E8mSSrsa} we have the evidently inequivalent rank-8 semisimple subalgebras $A_7{+}A_1$ and $E_7{+}A_1$. Omitting the larger summands, we obtain two subalgebras: $A_1\subset A_7{+}A_1\subset E_8$ and $A_1\subset E_7{+}A_1\subset E_8$. However, it turns out that these two different embeddings are in fact equivalent by $E_8$-conjugation\cite{rD-LieSub}, so that the centralizer of $A_1\subset E_8$ is always $E_7$; this $E_7$-centralizer subsumes the $A_7$ from the former subalgebra chain.

In turn, omitting $A_1$ from $A_7{+}A_1\subset E_8$ leaves the rank-7 subalgebra $A_7\subset A_7{+}A_1\subset E_8$ with $A_1$ the centralizer in $E_8$. Since $(A_7\subset A_7{+}A_1)\not\subset A_8\subset E_8$, the so-obtained subalgebra $A_7$ cannot be isomorphic to $A'_7$ in\eq{e:E8nSSrsa}. This identifies $A_7\subset A_7{+}A_1\not\subset A_8\subset E_8$ as Dynkin's $A''_7$ in\eq{e:E8nSSrsa}, since the first subalgebra pattern in\eq{e:nSSsa} and Dynkin's distinction of $A'_7$ imply that $(A_7\subset A'_7{+}K_1)\subset A_8\subset E_8$.

It turns out that the remaining isomorphic but inequivalently embedded pairs of four subalgebras,
\begin{equation}
  A_5{+}A_1,\qquad
  2A_3,\qquad
  A_3{+}2A_1\qquad
  \text{and}\qquad
  4A_1,
\end{equation}
are similarly distinguished by their (carefully traced) centralizers in $E_8$.
 The resulting 76 proper subgroups corresponding to these algebras (including the $U(1)^{8-r}$ abelian factor corresponding to the Cartan subalgebra) are listed in Table~\ref{atsg}.

\subsection{Maximal-Rank Regular Subgroups}
The preservation by the $\ZZ_N$-action\eq{shift} of the abelian factor $U(1)^{8-r}$ in $H_I\subset G$ renders the centralizer of $H_I\subset G$ equal to its center.

To see this, consider for example the distinct maximal regular subalgebras $A'_7\subset E_8$ and $A''_7+A_1\subset E_8$, where $A_7'\subset A_8\subset E_8$ whereas $A_7''\not\subset A_8\subset E_8$\cite{rD-LieSub}. Omitting the $A_1$ summand from the latter results in two inequivalently embedded $A_7$ subalgebras of $E_8$: the centralizer of $A_7'$ is 0, while the centralizer of $A''_7$ is $A_1$.

Passing to the corresponding compact Lie groups, we thus have the two inequivalently embedded $SU_8$ subgroups of $E_8$, shown here paired with their respective centralizers:
\begin{equation}
   \big\{\, SU'_8\subset E_8 \,,~ C_{E_8}(SU'_8)=U_1 \,\big\}
    \quad\textit{vs.}\quad
   \big\{\, SU''_8\subset E_8 \,,~ C_{E_8}(SU''_8)=SU_2 \,\big\}.
 \label{e:2SU8}
\end{equation}
The $\ZZ_N$-invariant subgroup\eqs{shift}{e:rkHI} of $E_8$ that contains an $SU_8$ factor is however $SU_8\times U_1$. In the case of $SU'_8$, this $\ZZ_N$-invariant $U_1$ factor is simply all of the centralizer\eq{e:2SU8}. For $SU''_8$ however, the $\ZZ_N$-invariant $U_1$ factor is a proper subgroup of the centralizer of $SU_8''$, $U_1\subset SU_2$, the centralizer of which is $C_{SU_2}(U_1)=\Ione$. Therefore, we obtain that
\begin{equation}
   \big\{\,(SU'_8\times U_1)\subset E_8 \,,~ C_{E_8}(SU'_8\times U_1)=\Ione \,\big\}
    \quad\textit{vs.}\quad
   \big\{\,(SU''_8\times U_1)\subset E_8 \,,~ C_{E_8}(SU''_8\times U_1)=\Ione \,\big\}.
 \label{e:2SU8U1}
\end{equation}
It then follows that the two $\ZZ_N$-invariant subgroups $SU_8\times U_1\subset E_8$, differing in the inequivalently embedded $SU_8$ factors, nevertheless have the same centralizer in $E_8$, equal to its center.
 The situation is similar for the other four subgroups,
 $SU_6\times SU_2\times U_1^{~2}$,
 $SU_4^{~2}\times U_1^{~2}$,
 $SU_4\times SU_2^{~2}\times U_1^{~3}$,
 $SU_2^{~4}\times U_1^{~4}$.

 As appropriate for the superstring and $M$-theory applications, which provided the original motivation for this analysis, we have herein not considered how the $\ZZ_N$-invariant subgroup $H_I$ and $\ZZ_N$ act on the $\ZZ_N$-{\em\/variant\/} complement of the adjoint representation---or any other $E_8$-representation.
 Also, as suggested by the Prof.~Milev, one could additionally partition the $\ZZ_N$-invariant roots of $E_8$ by the (integral) value of $\bm{v}{\cdot}\bm{p}$ in\eq{shift}.
 Such additional information should be able to clearly distinguish between the inequivalent embeddings of
\begin{equation}
  SU_8\times U_1,~~
  SU_6\times SU_2\times U_1^{~2},~~
  SU_4^{~2}\times U_1^{~2},~~
  SU_4\times SU_2^{~2}\times U_1^{~3},~~
  SU_2^{~4}\times U_1^{~4}~~\subset~ E_8,
 \label{doubles}
\end{equation}
and so provide a framework for a more detailed analysis then we had originally set out to explore.
 Along with a few other possible extensions of the present work as noted in the conclusions, we defer this line of inquiry to a separate effort, and for now remain content with listing only one copy of the subgroups\eq{doubles} in Table~\ref{atsg}, without any further distinction.

It is gratifying to note that the complete listing of maximal-rank regular subgroups of $E_8$ as given in Table~\ref{atsg} is also obtained by an iterative application of Tables~14 and 15 in Ref.\cite{rSsky}.

\par\bigskip
\providecommand{\href}[2]{#2}

\vfill
\end{document}